\documentclass[preprint,amsmath,amssymb,prfluids]{revtex4-2}

\usepackage{graphicx}
\usepackage{dcolumn}
\usepackage{bm}
\usepackage{hyperref}
\usepackage{color}
\usepackage{comment,amssymb,amsmath,mathtools}
\usepackage[normalem]{ulem}

\def\avk#1{\textcolor{black}{#1}}
\def\avkrev#1{\textcolor{black}{#1}}

\newcommand{\R}{\mathrm{Re}}
\newcommand{\tell}{\avkrev{\tilde{\ell}}}

\begin{document}

\preprint{APS/123-QED}

\title{Fluctuation-Induced Transitions in Anisotropic Two-Dimensional Turbulence}

\author{Lichuan Xu$^1$} \author{Adrian van Kan$^1$}\email{avankan@berkeley.edu}\author{Chang Liu$^2$} \author{Edgar Knobloch$^1$} 

\affiliation{$^1$ Department of Physics, University of California, Berkeley, CA 94720, USA}
\affiliation{$^2$ School of Mechanical, Aerospace, and Manufacturing Engineering, University of Connecticut, Storrs, CT 06269, USA}



\date{\today}

\begin{abstract}
Two-dimensional (2D) turbulence features an inverse energy cascade that produces large-scale flow structures such as hurricane-like large-scale vortices (LSVs) and unidirectional jets. We investigate the dynamics of such large-scale structures using extensive direct numerical simulations (DNS) of stochastically forced, viscously damped 2D turbulence within a periodic rectangular (Cartesian) domain $[0,L_x]\times[0,L_y]$. LSVs form and dominate the system when the domain aspect ratio $\delta = L_x/L_y \approx 1$, while unidirectional jets predominate at $\delta \gtrsim 1.1$. At intermediate values of $\delta$, both structures are metastable, and fluctuation-induced transitions between LSVs and jets are observed. Based on large-scale energy balance in the condensate, we derive and verify predictions for the dependence of the total kinetic energy and the flow polarity on the nondimensional control parameters. We further collect detailed statistics on the lifetimes of LSVs and jets from DNS runs of up to \avkrev{$10738$ viscous diffusive time units in length}. The distribution of the lifetimes is consistent with that of a memoryless Poisson process. 
\avkrev{The data are compatible with an exponential dependence of the mean lifetime on the aspect ratio $\delta$.} In addition, the mean lifetimes depend sensitively on the Reynolds number $\R$: as $\R$ increases, the energy gap between LSV (lower energy) and jet states (higher energy) \avk{arising from anisotropic dissipation} increases, leading to an increase in lifetimes that is \avkrev{approximately exponential} in $\R$ for both LSVs and jets. Similarly, as the \avkrev{ratio} of the forcing scale to the domain size increases, the transition rates increase sharply, confirming earlier findings. We investigate the transition dynamics in terms of kinetic energy, flow polarity, modal amplitude, and 2D phase-space diagrams, revealing that the transitions occur in two stages: in the initial stage, an efficient redistribution of kinetic energy by nonlinear triadic interactions facilitates a rapid transition from LSVs to jets and vice versa. In the second stage, the kinetic energy of the newly formed structure slowly adjusts to its associated (higher or lower) equilibrium value on a longer, viscous timescale, leading to a time delay that results in hysteretic transition behavior. Fluctuation-induced transitions may also occur between different numbers of jets. Our findings shed new light on the dynamics of coherent large-scale structures in anisotropic turbulence. 
\end{abstract}

\maketitle



\section{Introduction}

Turbulent flows are ubiquitous in nature. In planetary atmospheres and oceans, fluid motions are often restricted to being quasi-two-dimensional due to geometric confinement, rapid rotation, and stratification \cite{pedlosky2013geophysical,vallis2017atmospheric}. This provides a motivation for the study of highly idealized two-dimensional (2D) turbulence \cite{boffetta2012two}. A distinguishing feature of such 2D turbulence is the fact that it displays an inverse energy cascade \cite{kraichnan1967inertial}, leading in finite domains to the formation of large-scale coherent structures such as vortices and jets  \cite{chertkov2007dynamics,chan2012dynamics,frishman2017jets}. \avkrev{These flow structures, known as \textit{condensates} by analogy with Bose-Einstein condensates in a boson gas, were first observed in the pioneering experiments of Sommeria \cite{sommeria1986experimental}, and in numerical simulations in a square periodic domain by Smith et al. \citep{smith1993bose}. In physical space, condensates take different forms which resemble widely observed phenomena in geophysical flows, including hurricane-like vortices and zonal jets.} The formation of condensates is not limited to inertial 2D turbulence, as this process has been observed in models of living fluids \cite{linkmann2019phase,linkmann2020condensate,puggioni2022giant} and is also present in various highly anisotropic, three-dimensional (3D) flows in the presence of rotation, density stratification, thin-layer geometry and other effects \cite{rubio2014upscale,guervilly2014large,sS14,seshasayanan2018condensates,guervilly2017jets,julien2018impact,van2019condensates,alexakis2018cascades,alexakis2023quasi}.  
However, despite the prevalence of turbulent condensation phenomena across such a range of physical systems, the formation and dynamics of the associated turbulent large-scale flow structures remain incompletely understood.

 In \cite{bouchet2009random}, the authors studied stochastically forced 2D turbulence within a rectangular domain $[0,L_x]\times[0,L_y]$, with an aspect ratio $L_x/L_y\equiv \delta$, subject to periodic boundary conditions. The aspect ratio was varied, and two types of flow structures were observed. In a square domain, a pair of counter-rotating large-scale vortices (LSVs) forms, whose properties have been extensively studied in the past \cite{chan2012dynamics,laurie2014universal,frishman2018turbulence,svirsky2023two}. When the domain is elongated in one direction, jets oriented parallel to the short side of the domain emerge instead. The transition between these two disparate regimes is of particular interest. As the aspect ratio increases from $\delta=1$, the system traverses an interval of bistability, where the two different flow structures persist metastably (with a finite lifetime) until they are disrupted by random, fluctuation-induced transitions towards the competing structure. It has been confirmed theoretically, using tools from statistical mechanics \cite{bouchet2012statistical}, that coherent structures in the inviscid limit of the 2D Navier-Stokes equation (i.e. the 2D Euler equation) can indeed take the form of LSVs or jets. Statistical equilibria of 2D flows were also recently revisited in \cite{weichman2022statistical}.  \avkrev{Similar LSV and jet solutions were furthermore observed in anisotropically forced, direct statistical simulations of planar 2D turbulence in \cite{tobias2017direct}, and in solutions of the Euler equation on a sphere \cite{herbert2013additional,modin2020casimir}, although in the latter case ergodicity breaking may prevent the formation of a condensate in the absence of a forcing \cite{dritschel2015late,qi2014hyperviscosity}}. Moreover, LSVs and jets were reported in rotating convection within anisotropic domains \cite{julien2018impact,guervilly2017jets} and, more recently, in convection with misaligned gravity and rotation axes \cite{novi2019,barker2020,julien2022quasi,aE2023}. Dipoles and unidirectional jets were also found to play an important role in the inverse cascade of decaying 2D turbulence prior to condensation \cite{jimenez2020dipoles}.
 
 Bistability between qualitatively distinct metastable turbulent flow states (i.e. states whose lifetime is finite in the presence of fluctuations) is observed in a wide variety of physical systems, including body-forced three-dimensional (3D) turbulence within thin layers \cite{de2022bistability,van2019rare} and rotating turbulence \cite{yokoyama2017hysteretic}, where, in both cases, LSVs and small-scale 3D turbulence coexist over a range of the control parameter. A similar type of bistability was identified in rotating Rayleigh-Bénard convection  \cite{favier2019subcritical,de2022discontinuous}. Bistability was also found between states with and without super-rotation in a model of atmospheric dynamics \cite{herbert2020atmospheric} and in experiments on a wide range of turbulent flows \cite{weeks1997transitions,ravelet2004multistability,gayout2021rare}. However, when observation times are limited, multistability may not manifest itself through spontaneous transitions even in the presence of fluctuations, provided the latter are appropriately small. For instance, in \cite{favier2019subcritical}, no spontaneous transitions between 3D turbulence and LSVs were observed in the range of bistability in the simulations, despite the presence of turbulent fluctuations (whose amplitude was small compared to the energy gap between the two states). Similarly, while pronounced multistability was identified in instability-driven 2D turbulence between LSVs, vortex crystals, and mixtures of these \cite{van2022spontaneous,van2023vortex}, no spontaneous transitions were observed between these states, 
 although metastable vortex crystals have been observed in body-forced rotating 3D turbulence \cite{di2020phase}. Another example of bistability and rare transitions is found in barotropic (i.e. 2D) turbulence on the $\beta$-plane (an idealized model that includes a latitudinally varying Coriolis force, featuring zonal jets \cite{rhines1975waves}), where the number of zonal jets can change randomly at fixed parameters \cite{bouchet2019rare,cope2020dynamics,simonnet2021multistability}. Random transitions were also observed in \avkrev{2D} horizontally periodic convection with stress-free boundaries, between convection rolls and so-called windy states characterized by horizontal flow with strong vertical shear and suppressed convection 
 \cite{wang2020zonal,wang2023lifetimes}. Bistability and spontaneous transitions are also observed in the paths of the Gulf Stream and the Kuroshio current \cite{schmeits2001bimodal}\avkrev{, as well as in the thermohaline circulation \cite{timmermann2000noise}}. More generally, multistability is common in the climate system \cite{margazoglou2021dynamical}. 
 
 An important class of random transitions in turbulent flows involves spontaneous reversals. Such reversals are observed, for instance, in the polarity of the magnetic field of the Earth \cite{jacobs1994reversals} (where reversals occur randomly) and of the Sun \cite{charbonneau2014solar} (where they are part of the more regular 22 year solar cycle), a phenomenon reproduced qualitatively in laboratory dynamo experiments \cite{berhanu2007magnetic} and subsequent theoretical developments, cf. \cite{gallet2012reversals}. An additional example of a reversing flow is found at high altitudes in the Earth's atmosphere, where the quasi-biennial oscillation consists in reversals of the mean zonal wind approximately every two years \cite{baldwin2001quasi}, a phenomenon that has also been studied using laboratory analogs \cite{plumb1978instability,semin2018nonlinear}. In fact, reversals are common in convectively driven flows and have been the subject of detailed studies  \cite{sugiyama2010flow,ni2015reversals,wang2018mechanism,chen2019emergence,winchester2021zonal,liu2022staircase,liu2023fixed}. In addition, random reversals have also been observed and studied in confined 2D or quasi-2D turbulence, both in the lab \cite{sommeria1986experimental,michel2016bifurcations,fauve2017instabilities,pereira20191} and in numerical as well as theoretical work \cite{mishra2015dynamics, shukla2016statistical,dallas2020transitions,van2022geometric}. 

Beyond the specific case of turbulent flows, the study of fluctuation-induced transitions in random processes has a long history \cite{horsthemke1984noise}. A problem of particular importance concerning metastable states such as those discussed above is that of computing first-passage times, e.g., the first time at which a certain random process crosses a given threshold \cite{redner2001guide}. Problems of this type arise in a variety of areas from finance \cite{patie2004some}, chemistry \cite{almgren1984fluorescence} and biology \cite{burkitt2006review,chou2014first} \avkrev{to fluid flows \cite{riviere2024bubble}}. The classic first-passage problem concerns escape from a potential well due to small-amplitude noise, for which the mean first-passage time is given by the Eyring-Kramers formula \cite{langer1968theory,hanggi1990reaction,mel1991kramers}, which indicates that the mean first-passage time depends exponentially on the depth of the potential well. 
In the context of chemical kinetics, this result is associated with the name of Arrhenius \cite{arrhenius1889reaktionsgeschwindigkeit,laidler1984development,connors1990chemical}. A similar exponential dependence of the mean lifetime of metastable zonal jets on a (frictional) control parameter was recently identified in barotropic turbulence on the $\beta$-plane by Bouchet et al. in \cite{bouchet2019rare}, where the authors state that their results provide the first example of such an Arrhenius law in a turbulent flow. In general, the lifetimes of metastable states in turbulence in different systems have been found to depend on physical control parameters in a variety of ways, depending on the system under consideration. 
In body-forced thin-layer turbulence, the average lifetime of LSVs and 3D turbulence was found to change faster than exponentially with the nondimensional layer height \cite{van2019rare} and appears to diverge at a threshold height \cite{de2022bistability}. By contrast, in windy convection, the mean lifetime of metastable states was recently reported to exhibit a power-law dependence on the Rayleigh number with an exponent close to $4$ \cite{wang2023lifetimes}. 

Another important problem involving a highly non-trivial dependence of characteristic timescales on the physical control parameter is the transition to turbulence in pipe flow \cite{avila2023transition}. There, localized turbulent puffs systematically meet one of two fates: they can randomly split or decay, with each process having an associated time scale that follows a double-exponential function of the Reynolds number. Importantly, due to limitations of the available data, early works had misidentified this dependence as exponential, i.e. Arrhenius-like \cite{hof2006finite}, but more detailed measurements subsequently established the superexponential dependence as correct \cite{avila2011onset}, supported by theoretical arguments based on the insight that puff decay in pipe flow is triggered when the pointwise maximum turbulent intensity drops below some threshold \cite{goldenfeld2010extreme}. The intersection of the two double-exponential curves representing the time scales of turbulent puff decay and splitting, respectively, defines the critical Reynolds number for the transition to turbulence in a pipe. Similar superexponential dependence of turbulence lifetimes has also been reported for other transitional shear flows \cite{borrero2010transient, schneider2010transient,gome2020statistical,gome2022extreme}, disproving earlier claims of diverging lifetimes at a critical Reynolds number.

Against the backdrop of this rich literature on metastable states in turbulence and beyond, the present study focuses on fluctuation-induced transitions in stochastically forced 2D turbulence within anisotropic domains, \avkrev{arguably one of the simplest examples of a bistable turbulent flow}. In \cite{bouchet2009random}, which provides a major motivation for the work presented here, only a small number of transition events between LSVs and jets could be observed due to the relatively limited simulation time. For this reason, the statistical properties of this system have not yet been comprehensively analyzed. In particular, the distribution of first-passage times between LSVs and jets, and its dependence on the physical control parameters of the system are unknown, as are the phase space paths traversed by the system during transitions. Here, we go significantly beyond the scope of \cite{bouchet2009random} by conducting direct numerical simulations (DNS) over a very long time to obtain statistical information on the condensate states and the transition events, and analyze the impact of the three nondimensional control parameters (domain aspect ratio, Reynolds number, and nondimensional forcing scale).

The remainder of this paper is structured as follows. In Sec.~\ref{sec:setup}, we describe our simulation set-up and introduce the relevant nondimensional control parameters. Next, in Sec.~\ref{sec:structures}, we describe the large-scale structures arising in our system, introduce the quantitative measures we use to identify them, and formulate a simple model based on large-scale energy balance. In Sec.~\ref{sec:random}, we present the phenomenology of the random transitions between LSVs and jets and discuss the impact of all nondimensional control parameters on the transitions. In Sec.~\ref{sec:transitions}, we examine the phase space trajectories traversed by the system during transitions, while Section~\ref{sec:jets} examines 
spontaneous transitions between different numbers of jets. The paper concludes in Sec.~\ref{sec:conclusions} with a discussion of  our results and outlines remaining open questions. 


\section{Simulation Set-up\label{sec:setup}}
\subsection{Equations and important quantities}
\begin{figure}
\includegraphics[width = 0.45\textwidth]{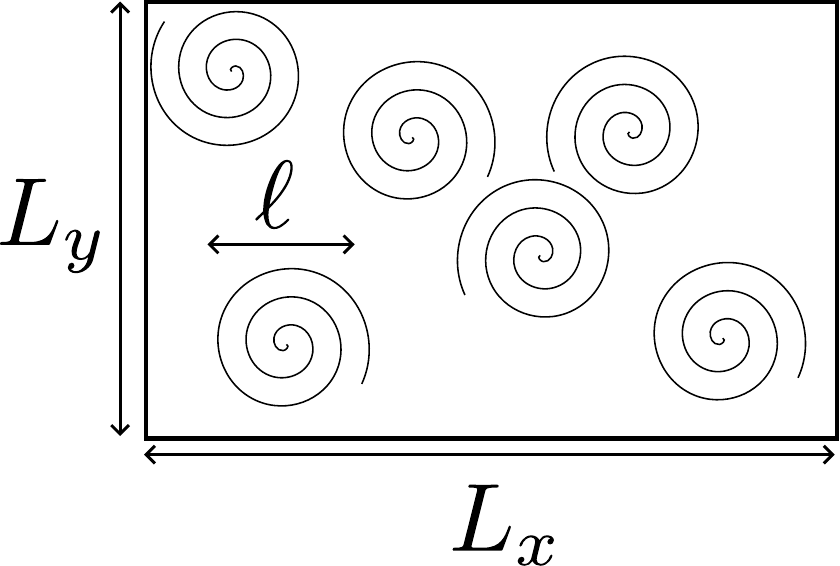}
\caption{An illustration of the domain with side lengths $L_x$ and $L_y$ ($L_x>L_y$), and random forcing acting on the length scale $\ell$.}
\label{fig:domain_sketch}
\end{figure}

We study incompressible flow of a constant density fluid within a planar rectangular domain $[0,L_x]\times[0,L_y]$, illustrated in Fig.~\ref{fig:domain_sketch}. The fluid motion is driven by a stochastic force, and the domain is subject to periodic boundary conditions. The flow obeys the 2D Navier-Stokes equation
\begin{equation}
    \partial_t \mathbf{u}+\mathbf{u} \cdot \nabla \mathbf{u}=-(1 / \rho) \nabla p + \nu\nabla^2\mathbf{u} + \mathbf{f}, \quad \nabla \cdot \mathbf{u} = 0,
    \label{eq:NSE}
\end{equation}

\noindent where $\mathbf{u}=u\mathbf{\hat{x}} + v\mathbf{\hat{y}}$ is the 2D velocity field, $\rho$ is the constant density, $p$ is the pressure, $\nu$ is the kinematic viscosity and $\mathbf{f}$ is the driving force. \avkrev{The force $\mathbf{f}$ is chosen to have zero spatial mean, to be random and white in time}, acting within an isotropic forcing band centered on a typical forcing scale $\ell$ \avkrev{(defined as the arithmetic mean of the largest and smallest length scales in the forcing band)} and injecting energy at a rate $\epsilon$. \avkrev{In Appendix \ref{sec:forcing_details}, we describe in detail how the width of the (narrow) forcing band is selected.} The forcing spectrum is flat over the forcing band and the randomness consists in uniformly distributed phases of the complex Fourier modes. To simplify Eq.~\eqref{eq:NSE}, we introduce the stream function $\psi (x, y, t)$ such that $\mathbf{u}=\left(\partial_y \psi,-\partial_x \psi\right)$, satisfying the incompressibility constraint $\nabla \cdot \mathbf{u} = 0$. We note that in our set-up, following \cite{bouchet2009random}, the only mechanism for saturation of large-scale condensates in Eq.~(\ref{eq:NSE}) is viscous dissipation, which decreases in strength as the length scale increases. Many studies of 2D turbulence therefore include additional saturation mechanisms, such as Rayleigh friction \cite{boffetta2012two,frishman2017jets}. In the interest of simplicity we retain here only viscous dissipation in order to reduce the number of model parameters. \avkrev{It is known that the condensate properties, including the qualitative spatial profile of the large-scale flow, depend explicitly on the choice of saturation mechanism in 2D turbulence (specifically, whether the condensate saturates by viscosity or Rayleigh friction \cite{frishman2017culmination,doludenko2021coherent}}). For this reason the dynamics of large-scale structures described below are not expected to follow the results of \cite{frishman2017jets}, \avkrev{where Rayleigh friction is considered.}

The system is characterized by three \avkrev{a priori} nondimensional control parameters: the aspect ratio $\delta$, the ratio $\tilde{\ell}$ between forcing scale and domain size, and the forcing-scale Reynolds number $\R$, defined as
\begin{equation}
    \delta \equiv L_x/L_y, \quad \avkrev{\tilde{\ell}\equiv\ell/L_y}, \quad \mathrm{Re}\equiv\epsilon^{1/3}\ell^{4/3}/\nu,
    \label{eq:nd_params}
\end{equation}
respectively. 
The average kinetic energy density of the flow is given by
\begin{equation}
E \equiv \frac{1}{2} \langle|\mathbf{u}|^2\rangle = \frac{1}{2}\left( \langle u^2\rangle + \langle v^2 \rangle\right),\label{eq:Ekin}
\end{equation}
where $\langle \cdot \rangle$ is the domain average. A typical velocity scale dictated by the forcing is $u_f = (\epsilon\ell)^{1/3} $, giving the forcing energy scale  
\begin{equation}
    E_f =\frac{1}{2}(\epsilon\ell)^{2/3}.
    \label{eq:Ef}
\end{equation} 
In addition, an important time scale characterizing the evolution of the system is the \avkrev{time it takes for momentum to diffuse across the entire domain (for specificity, in the $y$ direction)}
\begin{equation}
    t_\nu = \frac{L_y^2}{4\pi^2 \nu}.
    \label{eq:tnu}
\end{equation}

\avkrev{We emphasize that $t_\nu$ is a natural choice of time scale for analyzing the long-term behavior of this system. In the presence of a strong condensate, the shortest characteristic time in the system is the large-scale turnover time $t_{ls}=L_y/U_{ls}$. From large-scale energy balance, it follows that $E\approx \frac{1}{2}U^2_{ls}\propto E_f \mathrm{Re}/\tilde{\ell}^2$, cf.  Eq.~(\ref{eq:theo_pred_E_nondim}). This implies that 
\begin{equation}t_{ls}\propto \underbrace{(\ell^2/\epsilon)^{1/3}}_{\equiv t_f}\frac{1}{\sqrt{\mathrm{Re}}},
\end{equation}
    in terms of the time scale of enstrophy injection by the forcing $t_f=(\ell^2/\epsilon)^{1/3}$, which is the second important time scale in the problem, and which also coincides with the eddy turnover time in the forward enstrophy cascade. The slowest a priori characteristic time scale in the system is $t_\nu\propto t_f \,\mathrm{Re}$. In the limit of large $\mathrm{Re}$, these time scales are well separated,
    \begin{equation}
    t_{ls} \ll t_f \ll t_\nu.   
    \end{equation}
    Hence, $t_\nu$ is the key time scale for analysing the long-time evolution of the system.}

 In the following, we denote the nondimensional time measured in viscous units by $\widetilde{t} = t/t_\nu$ and the nondimensional energy measured in units of the forcing energy scale by $\widetilde{E}=E/E_f$. 
We integrate Eq.~\eqref{eq:NSE} subject to periodic boundary conditions using the Fourier pseudo-spectral solver GHOST \cite{mininni2011hybrid}, where the 2/3 rule is adopted for dealiasing and a semi-implicit fourth-order Runge-Kutta scheme is used for time integration. All simulations are initialized with small amplitude random initial conditions.

{Table \ref{tab:summary_Runs} shows a summary of all simulations performed in this study. All the runs were performed with $L_x \geq L_y$, such that the aspect ratio $\delta \geq 1$. \avkrev{The runs are grouped into several sets: in sets $\mathrm{D1}$ and $\mathrm{D2}$, the aspect ratio $\delta$ is varied over the relatively narrow region of bistability between LSVs and jets $1\leq \delta \leq 1.1$ for fixed $\R=459$ or $\R=344$, respectively, and fixed nondimensional forcing scale. In set $\mathrm{R}$ the Reynolds number is varied between $\R=275$ and $\R=688$, while keeping $\delta$ and $\tell$ fixed, and in set $\mathrm{L}$ the nondimensional forcing scale is varied between $\tilde{\ell}=0.173$ and $\tilde{\ell}=0.49$. Run $\mathrm{T}$ exhibits tristability between jets in either direction and the LSV state in a square domain for $\tell=1/\sqrt{5}$, while run $\mathrm{J}$ shows a transition between distinct numbers of jets in a highly elongated domain with $\delta>3$. All runs analysed here are long compared to the viscous time scale $t_\nu$, with the shortest runtime being $60\,t_\nu$ and the longest runtime being \avkrev{$10738\,t_\nu$}. We stress that the longest runs reported here are more than two orders of magnitude longer than the simulations reported in \cite{bouchet2009random}, which only reached $\lesssim 50\,t_\nu$. In order to be able to integrate for such long times, a requirement of the very long characteristic time scales in the problem, we adopt moderate spatial resolutions of $128 \times 128$ or $256\times 256$ grid points. The resolution is held fixed for any given Reynolds number as we vary $\delta$, i.e., we consider 
\avkrev{grids with slightly differing spacing in the $x$ and $y$ directions} in the interest of efficiency gains in the Fourier transforms.} 
\begin{table}
    \centering
    \begin{tabular}{|c|c|c|c|c|c|c|}
        \hline Set& {$\R$}& $n_x \times n_y$ & $\delta $ & $\tilde{\ell}$ & \# runs & maximum runtime in $t_\nu$\\ \hline
        D1 & 459 & $256\times 256$ & 1---1.1 & 0.34 & 16 & 2800 \\\hline 
        D2 & 344 & $128\times 128$& 1.045---1.1 & 0.34 & 12 & 10738 \\\hline 
        R & 275---688 & $128\times 128$, $256\times 256$ & 1.07 & 0.34 & 7 & 7219 \\\hline 
        L & 69 & $256\times 256$ & 1.07 & 0.173---0.491 & 8 & 670 \\\hline 
        T & 550 & $256\times 256$ & 1 & $=1/\sqrt{5}\, (\approx 0.45)$ & 1 & 230 \\\hline 
        J & 55 & $128\times 64$ & $3.1125$ & $0.53$ & 1 & $60$ \\\hline 
    \end{tabular}
    \caption{Summary of the runs described in this study. Columns show run label, Reynolds number, the number of grid points in the $x$ and $y$ directions, the aspect ratio $\delta$, the \avkrev{nondimensional forcing scale $\tilde{\ell}$}, the number of runs performed within each set of simulations and the maximum runtime within that set. }
    \label{tab:summary_Runs}
\end{table}

 \avkrev{We ensure that all runs reported here are well resolved by verifying that the \avkrev{enstrophy} dissipation spectrum $D_\nu(k) \equiv 2\nu k^4 E(k)$ decays toward the truncation scale, where $E(k)$ is the energy spectrum, binned in wave number intervals of $2\pi/L_x$, namely}
\begin{equation}
    \avkrev{E(k) = \frac{1}{2}\sum_{\stackrel{\mathbf{p}}{k-\pi/L_x\leq |\mathbf{p}|< k+\pi/L_x }} |\hat{\mathbf{u}}(\mathbf{p})|^2},\label{eq:energy_spectrum}
\end{equation}
\avkrev{where the hat indicates a Fourier transform. The enstrophy dissipation and energy spectra are shown in Fig.~\ref{fig:spectra} for a sample run from set D1.}

\begin{figure}
    \centering
    \hspace{0cm} (a) \hspace{8cm} (b)    \includegraphics[width=0.98\textwidth]{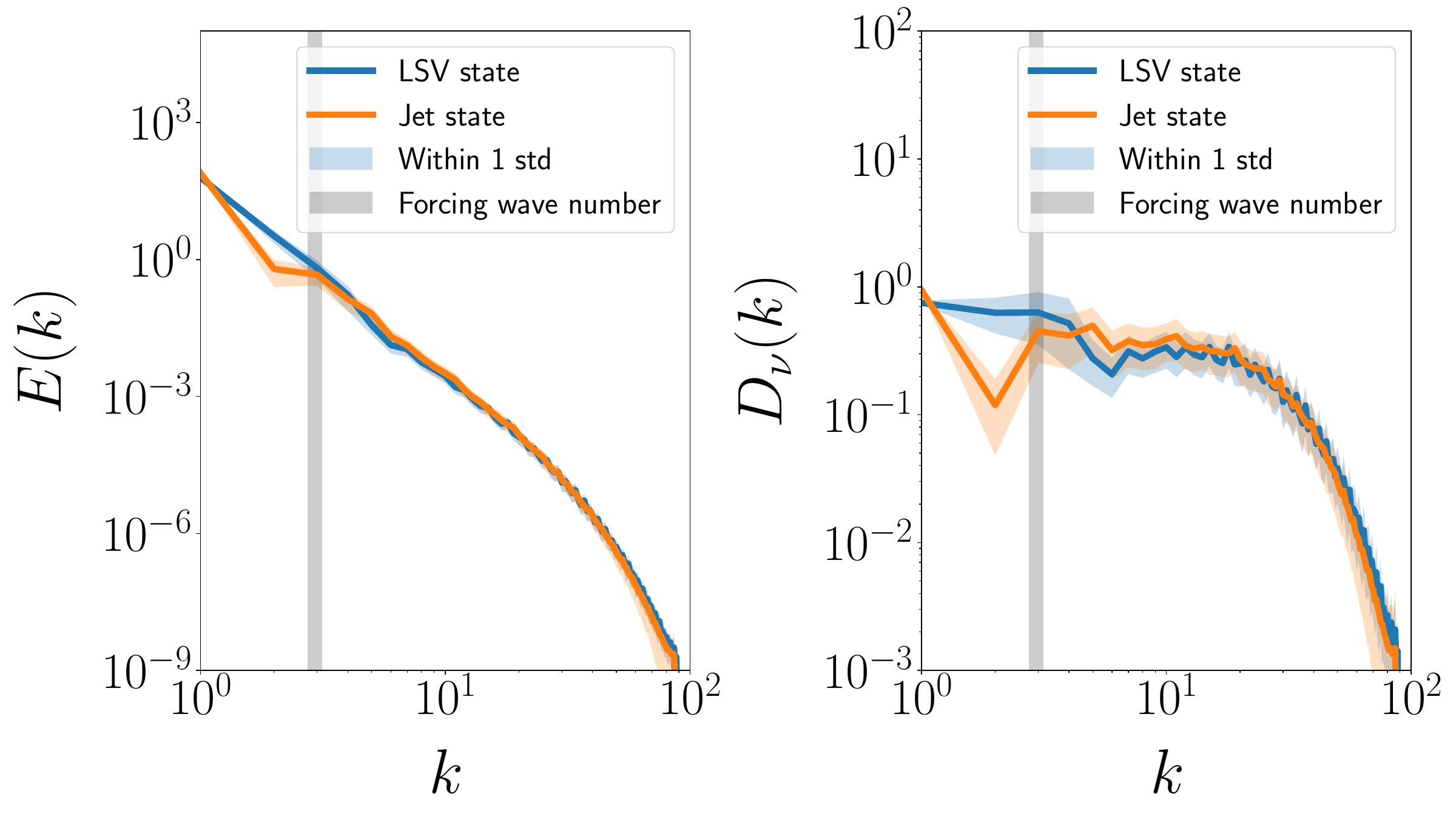}
    \caption{ \avkrev{Log-log plots of (a) energy spectrum and (b) enstrophy dissipation spectrum in the LSV state (blue line) and in the jet state (orange line). Gray vertical shaded region indicates the forcing region. Blue/orange shaded regions indicate one standard deviation (std) about the mean spectrum (computed based on $310$ outputs in the statistically stationary state for the LSV state, and $458$ outputs for the jet state. Data shown were obtained at $\delta=1.07$, $\R=459$, $\tell=0.34$ (set D1). The large-scale maximum in panel (a) reveals the presence of the condensate, while panel (b) highlights well-resolvedness of our simulations by the fact that the enstrophy dissipation spectrum decays towards the grid scale.}}
    \label{fig:spectra}
\end{figure}
\avkrev{In Appendix \ref{sec:remap_square}, we describe a remapping of the above setup to the square domain, which reveals effective anisotropic diffusion and forcing. Similar features are also found in the barotropic vorticity equation for quasi-geostrophic convection on a tilted $f$-plane characterized by misaligned rotation and gravity axes \cite{julien2022quasi,aE2023}, where transitions between LSVs and jets were observed at finite inclination angles, not unlike those which we explore further below. We stress that in \cite{julien2022quasi,aE2023} the anisotropy is not geometrical but rather derives from the misalignment between gravity and the rotation axis.}

\begin{figure}[!htbp]
\includegraphics[width = 0.75\textwidth]{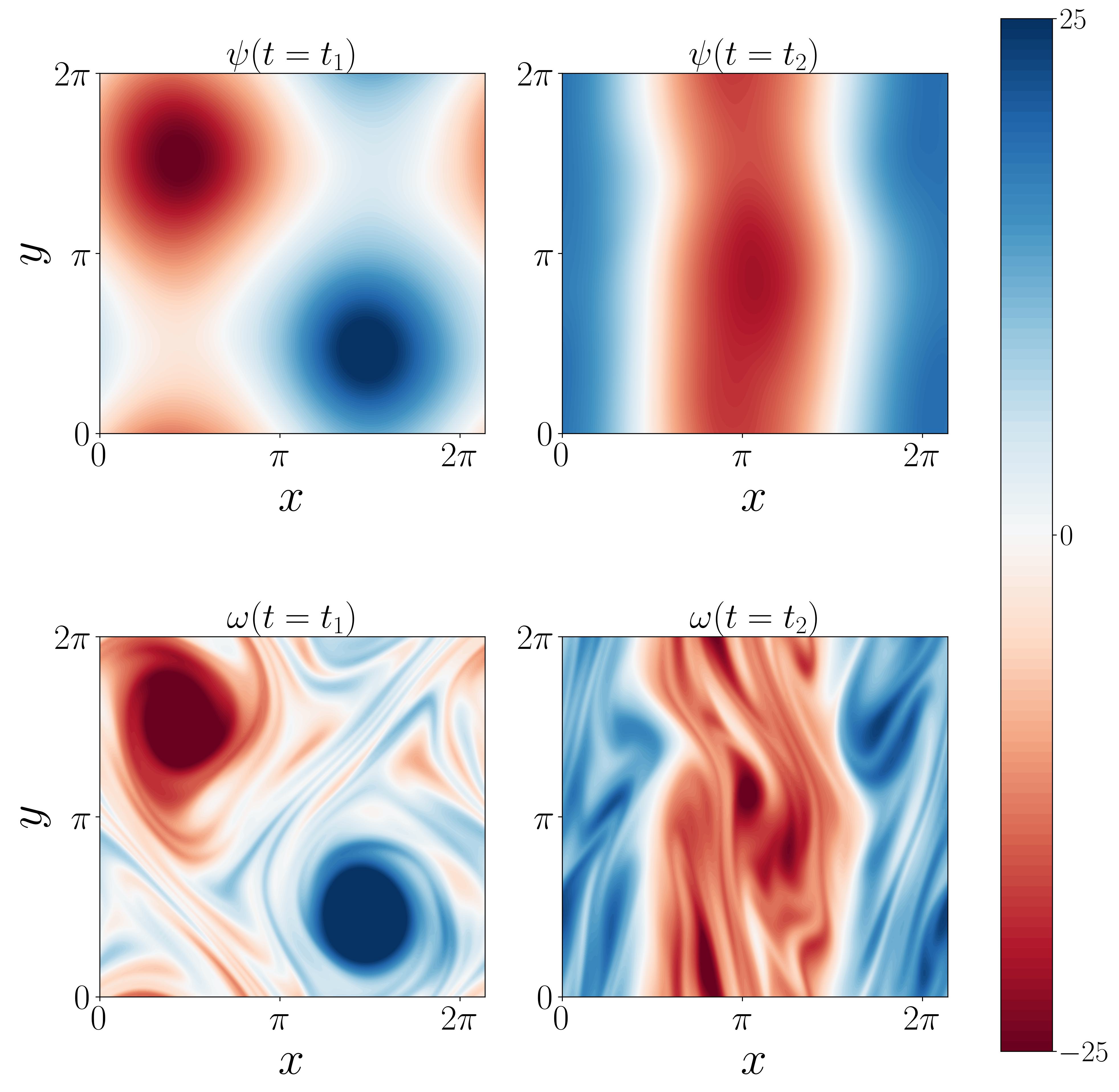}
\caption{Large-scale structures from a run from set D1 in a domain with aspect ratio $\delta=1.07$ at $\R=459$, $\tilde{\ell}=0.34$ (resolution $n_x=n_y=256$ grid points) at two times $t=t_1$, $t=t_2$ from a long simulation, visualized via the stream function $\psi(x,y)$ (top row) and vorticity $\omega(x,y)$ (bottom row). Left column: pair of LSVs present at $t=t_1$. Right column: unidirectional jet state present at $t=t_2$.}
\label{fig:LSVs_and_jets}
\end{figure}
\section{Bistable dynamics and large-scale structures \label{sec:structures}}
\subsection{Flow structures and proxies for their identification}
Energy injected by the random force $\mathbf{f}$ in the forcing band centered on the scale $\ell<L_y\leq L_x$ is transferred via the inverse cascade to larger scales. The viscous dissipation at large scales is weaker than at small scales for a given velocity amplitude, allowing the energy to accumulate at the scale of the domain. As the resulting condensate grows in amplitude, the weakness of large-scale viscous dissipation is compensated by increased velocity amplitude, thereby balancing the injection rate of energy and producing a statistically steady state where the total energy in the system fluctuates about a constant value.

In physical space, this condensation process results in the formation of large-scale coherent structures, as shown in Fig.~\ref{fig:LSVs_and_jets}.   Which type of structure form depends on the geometry of the domain, specifically on the aspect ratio $\delta$. In a square domain ($\delta=1$), large-scale vortices (LSVs) form, while in a highly rectangular domain ($\delta \gg 1$), unidirectional jets along the short side of the domain are found. Since the domain average of the vorticity $ \omega= -\nabla^2 \psi$, expressed in terms of the stream function $\psi$, vanishes exactly due to the periodic boundary conditions, LSVs and jets always occur in pairs. That is, the LSVs in the periodic domain always appear as a pair of counter-rotating vortices, while jet states \avkrev{invariably} consist of an even number of parallel bands with velocity of alternating sign. 


{We further note that in the infinite plane, Eq.~(\ref{eq:NSE}) is invariant under arbitrary rotations, i.e., the problem is isotropic. In the periodic square, by contrast, this symmetry is reduced to invariance under rotations by $90^\circ$. Hence, the set of solutions in the square domain is also invariant under this transformation. This symmetry is broken when the domain becomes rectangular, which is key for the transition from vortices (which are invariant under rotations by $90^\circ$) and unidirectional jets (which are not).}

The acquisition of statistical data requires that we quantitatively distinguish between the two large-scale structures. Thus we define the \textit{polarity} $m$ of the flow  based on the difference in the fraction of kinetic energy contained within the $x$ and $y$ components of the velocity,

\begin{equation}
    m \equiv \frac{\langle v^2\rangle - \langle u^2 \rangle}{\langle u^2\rangle + \langle v^2 \rangle}, \label{eq:polarisation}
\end{equation}

\noindent where $\langle \cdot \rangle$ denotes the spatial average over the domain. When a pair of LSVs is present in the domain, the kinetic energies in the $x$ and $y$ directions are approximately equal ($\langle v^2\rangle \approx \langle u^2 \rangle$), implying $m \approx 0$.  When jets in the $y$ direction are present, the flow is primarily oriented along the $y$ direction ($\langle v^2\rangle \gg \langle u^2 \rangle$), and $m \approx 1$. Similarly, jets in the $x$ direction correspond to $m \approx -1$. The quantity $m$ has been used to distinguish jets and LSVs in rapidly rotating convection within anisotropic domains \cite{guervilly2017jets} and an analogous measure was introduced in dynamo theory to distinguish between magnetic fields with dipolar and quadrupolar symmetry \cite{knobloch1998modulation}.

\begin{figure}[!htbp]
\includegraphics[width = 0.75\textwidth]{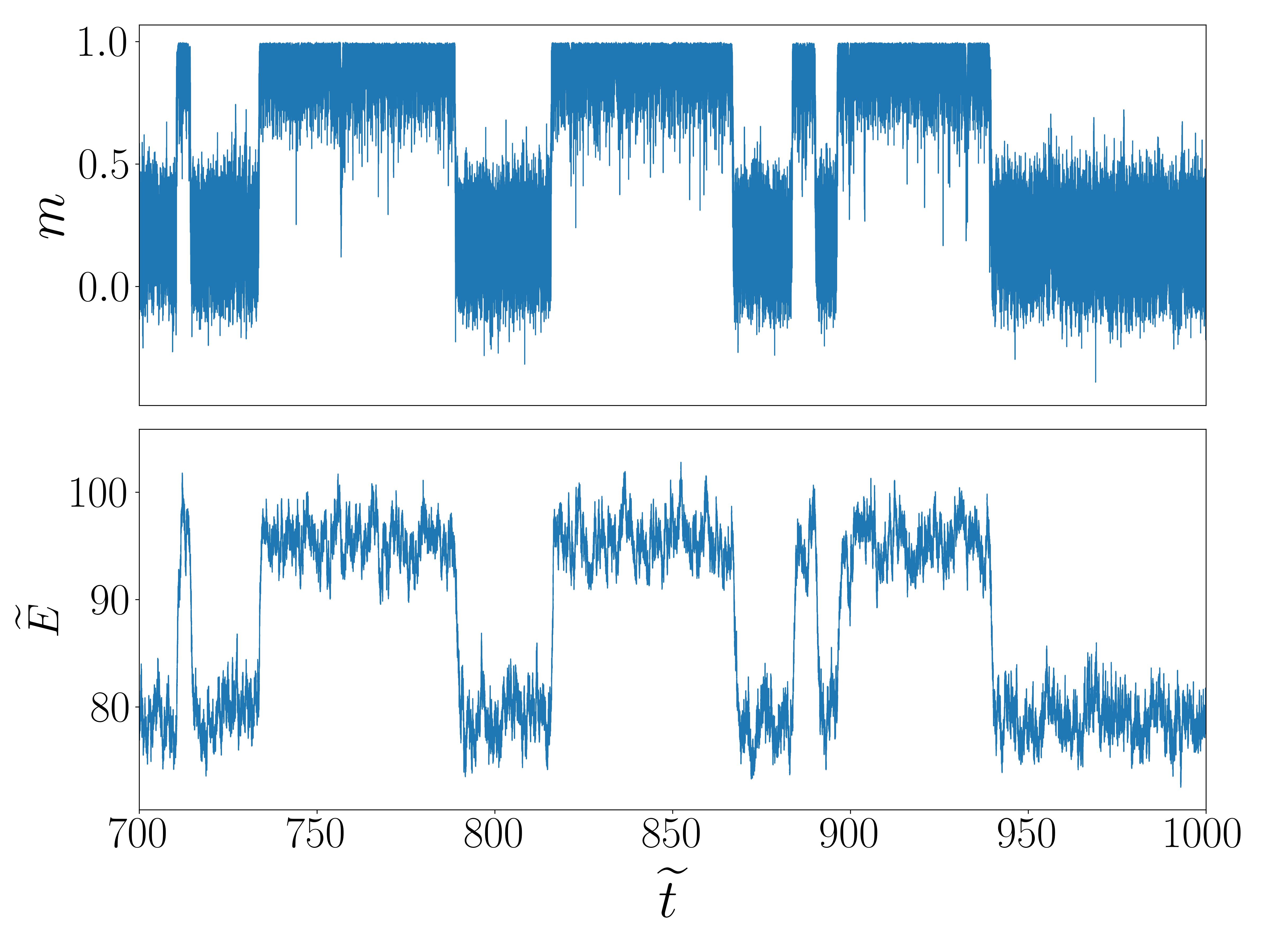}
\caption{The evolution of the polarity $m$ and the nondimensional energy $\widetilde{E}=E/E_f$ in units of the forcing energy $E_f$, defined in Eqs.~(\ref{eq:Ekin}) and (\ref{eq:polarisation}), during random, fluctuation-induced transitions between LSVs and jets. The $x$ axis shows nondimensional time measured in viscous units. Here $\delta = 1.065$ and $\mathrm{Re} = 459$ (set D1). When jets are present, $m\approx 1$, and $\widetilde{E}$ is at a high value. By contrast, when LSVs are present, $m$ is closer to $0$, and $\widetilde{E}$ is at a lower value. }
\label{fig:scalars_time_series}
\end{figure}

Figure \ref{fig:scalars_time_series} shows a section of the time series of both the polarity $m$ and the nondimensional kinetic energy $\widetilde{E}=E/E_f$, as defined in Eqs.~(\ref{eq:Ekin}) and (\ref{eq:polarisation}), at an aspect ratio of $\delta = 1.065$ and moderate Reynolds number $\R=459$. The flow spontaneously transitions between unidirectional jets in the (shorter) $y$ direction, characterized by $m\approx 1$, and LSVs, where $m$ is closer to zero (\avkrev{$m$ differs slightly from zero in the LSV state since vortices in the anisotropic domain are slightly elongated}). The kinetic energy of the vortex state is seen to be lower than that of the jet, an observation we discuss further below.

\subsection{Energy balance of large-scale condensates}
\label{sec:energy_balance_theory}
The parameter dependence of the polarity and the kinetic energy can be estimated by considering the large-scale energy balance in the statistically steady state. 
For the analysis of the spectral energy distribution, \avkrev{we introduce the combined kinetic energy in wave vectors 
$\pm\mathbf{k}_{a,b}=\pm\left(aQ_x,bQ_y\right)$, with $Q_x=2\pi/L_x$, $Q_y=2\pi/L_y$, denoted by} 
\begin{equation}
    E_{a,b} \equiv 
    \frac{1}{2} \left(\left|\hat{\mathbf{u}}\left(\mathbf{k}_{a,b}\right)\right|^2 + \left|\hat{\mathbf{u}}\left(-\mathbf{k}_{a,b}\right)\right|^2\right) = \left|\hat{\mathbf{u}}\left(\mathbf{k}_{a,b}\right)\right|^2, 
    \label{eq:Eab}
\end{equation}
\noindent where $\hat{\mathbf{u}}\left(\mathbf{k}\right)$ is the Fourier transform of the velocity with wave number $\mathbf{k}$. Equation~(\ref{eq:Eab}) takes into account the fact that the velocity field is real, i.e., that $\hat{\mathbf{u}}(\mathbf{k}) = \hat{\mathbf{u}}^*(-\mathbf{k})$.

In a nearly square domain with $\delta \gtrsim 1$, the two smallest wave numbers in the system are $Q_x$ and $Q_y$, which are associated with the wave vectors $\mathbf{k}_{\pm 1,0} = (\pm Q_x,0)$ and $\mathbf{k}_{0,\pm1}=(0,\pm Q_y)$, respectively. Energy is injected into the system at a rate $\epsilon$ within the forcing band centered on the length scale $\ell$, and transferred by the inverse cascade to the largest scales. A fraction $\alpha_{1,0}\,\epsilon$ of the injected energy is transferred to $\mathbf{k}_{\pm 1,0}$ and another fraction $\alpha_{0, 1}\,\epsilon$ is transferred to $\mathbf{k}_{0,\pm1}$, \avkrev{where $\alpha_{1,0}+\alpha_{0,1}\lesssim  1$ since a (small) part of the energy transferred upscale from the forcing will be dissipated at scales other than the largest ones, and one expects $\alpha_{1,0}+\alpha_{0,1}$ to increase with $\R$.} 
In steady state, the nonlinear transfer of energy from the forcing scales to the large-scale modes is balanced by the rate of dissipation, formally
\begin{align}
 \frac{\avkrev{2\nu  Q_y^2E_{1,0} }}{\delta^2} = \alpha_{1,0} \epsilon,  \quad \quad 
 \avkrev{2\nu Q_y^2  E_{0,1}} = \alpha_{0,1} \epsilon.
    \label{eq:energy_balance_steady_state}
\end{align}
For any given wave number $\mathbf{k}$, we can directly \avkrev{measure} 
the rate of energy received 
from the nonlinear transfer rate of energy to this mode, namely
\begin{equation}
    T(\mathbf{k}) = -2 \, \mathrm{Re}\left[\hat{\mathbf{u}}^*(\mathbf{k})\cdot \widehat{\mathbf{N}}(\mathbf{k}) \right], 
    \label{eq:Tab}
\end{equation}
where hats indicate Fourier transforms and $\mathbf{N} = (\mathbf{u}\cdot \nabla )\mathbf{u}$ is the advection term. 
Time-averaging this nonlinear transfer rate and adding up the contributions from $\pm \mathbf{k}_{a,b}$ yields $\alpha_{1,0}\,\epsilon$ and $\alpha_{0,1}\,\epsilon$, respectively, 
\begin{equation}
    \alpha_{a,b}\,\epsilon = \overline{T(\mathbf{k}_{a,b}) + T(-\mathbf{k}_{a,b})},
    \label{eq:transfer_rate_averaged}
\end{equation}
in terms of the time average $\overline{(\cdot)}$ computed over several lifetimes of a given structure. The particular cases $(a,b)=(1,0)$ and $(a,b)=(0,1)$ will be of relevance below.
\begin{figure}
    \centering
    \hspace{0cm} (a) \hspace{8cm} (b)
    \includegraphics[width = \textwidth]{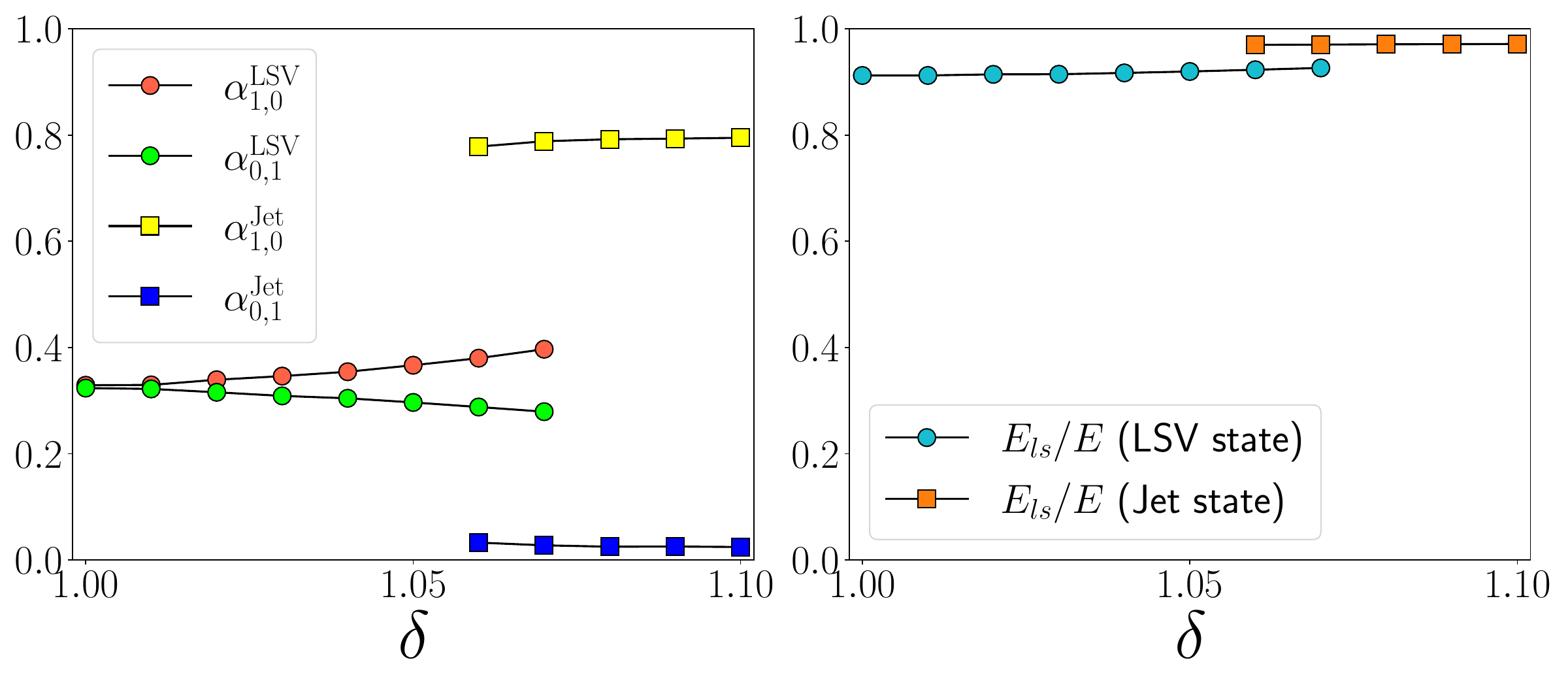}
    \caption{(a) Fraction of the injected energy fluxes $\alpha_{1,0}$, $\alpha_{0,1}$ maintaining the large-scale energies $E_{1,0}$ and $E_{0,1}$ against dissipation, cf. Eqs.~(\ref{eq:energy_balance_steady_state}), measured directly from the nonlinear transfer rate according to Eq.~(\ref{eq:transfer_rate_averaged}) \avkrev{based on runs in set D1}. As expected, in the LSV state, $\alpha_{1,0}=\alpha_{0,1}$ at $\delta=1$, with a slight dominance for the gravest (largest-scale) mode as $\delta $ increases, manifested by an increase in $\alpha_{1,0}$ and a decrease in $\alpha_{0,1}$. In the jet state, $\alpha_{0,1}\ll \alpha_{1,0}$ as expected. 
    A small fraction of kinetic energy is lost during the transfer from the forcing scales to the largest scales, so that $\alpha_{1,0}<1$ in the jet state. (b) Fraction of total kinetic energy $E$ residing in the large-scale contributions $E_{ls}=E_{1,0}+E_{0,1}$. Owing to the presence of the condensate, this fraction is close to one: in the LSV state  $E_{ls}/E\gtrsim 91\%$ and in the jet state $E_{ls}/E\gtrsim 96\%$.} 
    \label{fig:transfer_rates}
\end{figure}

Figure \ref{fig:transfer_rates}(a) shows the result of a direct measurement of $\alpha_{1,0}$ and $\alpha_{0,1}$ from set D1 of our DNS at $\R=459$ as a function of $\delta$. In the jet state, one finds, as expected,  that $\alpha_{0,1}\ll 1$ while $\alpha_{1,0}$ is close to one: the inverse cascade exclusively feeds the mode $\mathbf{k}_{1,0}$ in the $x$ direction. Moreover, in the LSV state, one finds comparable values for $\alpha_{1,0}$ and $\alpha_{0,1}$, with equality $\alpha_{1,0} = \alpha_{0,1}$ at $\delta=1$. As the aspect ratio increases from $\delta=1$, the difference between $\alpha_{1,0}$ and $\alpha_{0,1}$ increases, with a decrease in $\alpha_{1,0}$ and an increase in $\alpha_{0,1}$. 
\avkrev{Despite some amount of 
dissipation at smaller scales, the kinetic energy is clearly dominated by the large-scale contributions $E_{ls}=E_{1,0}+E_{0,1}$, with $E_{ls}/E\gtrsim91\%$ of the total energy residing within the large-scale modes in the LSV state, as shown in Fig.~\ref{fig:transfer_rates}(b). In the jet state, this fraction increases to $E_{ls}/E\gtrsim 96\%$ of the total energy contained within the largest-scale modes, cf. Fig.~\ref{fig:transfer_rates}(b).}

Since the kinetic energy $E$ is dominated by the largest scales in the presence of a strong condensate, Eq.(\ref{eq:energy_balance_steady_state}) implies an approximate expression
\begin{align}
    E \approx E_{1,0}+E_{0,1}  = \frac{\epsilon }{2 Q_y^2 \nu}\left(\alpha_{1,0} \delta^2 + \alpha_{0,1}\right).
    \label{eq:theo_pred_E}
\end{align}
In nondimensional terms, this amounts to
\begin{equation}
    \widetilde{E}=\frac{E}{E_f} \approx \frac{Re}{4\pi^2\tilde{\ell}^2} (\alpha_{1,0} \delta^2 + \alpha_{0,1}),
    \label{eq:theo_pred_E_nondim}
\end{equation}
where the forcing energy scale $E_f$ is defined in Eq.~(\ref{eq:Ef}). Similarly, from Eq.~(\ref{eq:energy_balance_steady_state}), we can also estimate the polarity of the flow as
\begin{align}
    m \approx \frac{ E_{1,0}- E_{0,1}}{E_{ls} }  = \frac{\alpha_{1,0} \delta^2 -\alpha_{0,1}}{\alpha_{1,0} \delta^2 + \alpha_{0,1}}. \label{eq:theo_pred_m}
\end{align}
Thus in the jet state, where $\alpha_{0,1}\ll 1$, the polarity $m \approx 1$, while in the LSV state the polarity $m$ is expected to vanish at $\delta=0$ (since $\alpha_{1,0}\approx \alpha_{0,1}$) and to become positive at $\delta>1$, since $\alpha_{1,0}>\alpha_{0,1}$ at $\delta>1$ (cf. Fig.~\ref{fig:transfer_rates}(a)). Next, we present detailed measurements of $m$ and $\widetilde{E}$ from DNS and confront the above predictions with the DNS results. 

\begin{figure}[!htbp]
\includegraphics[width = 0.75\textwidth]{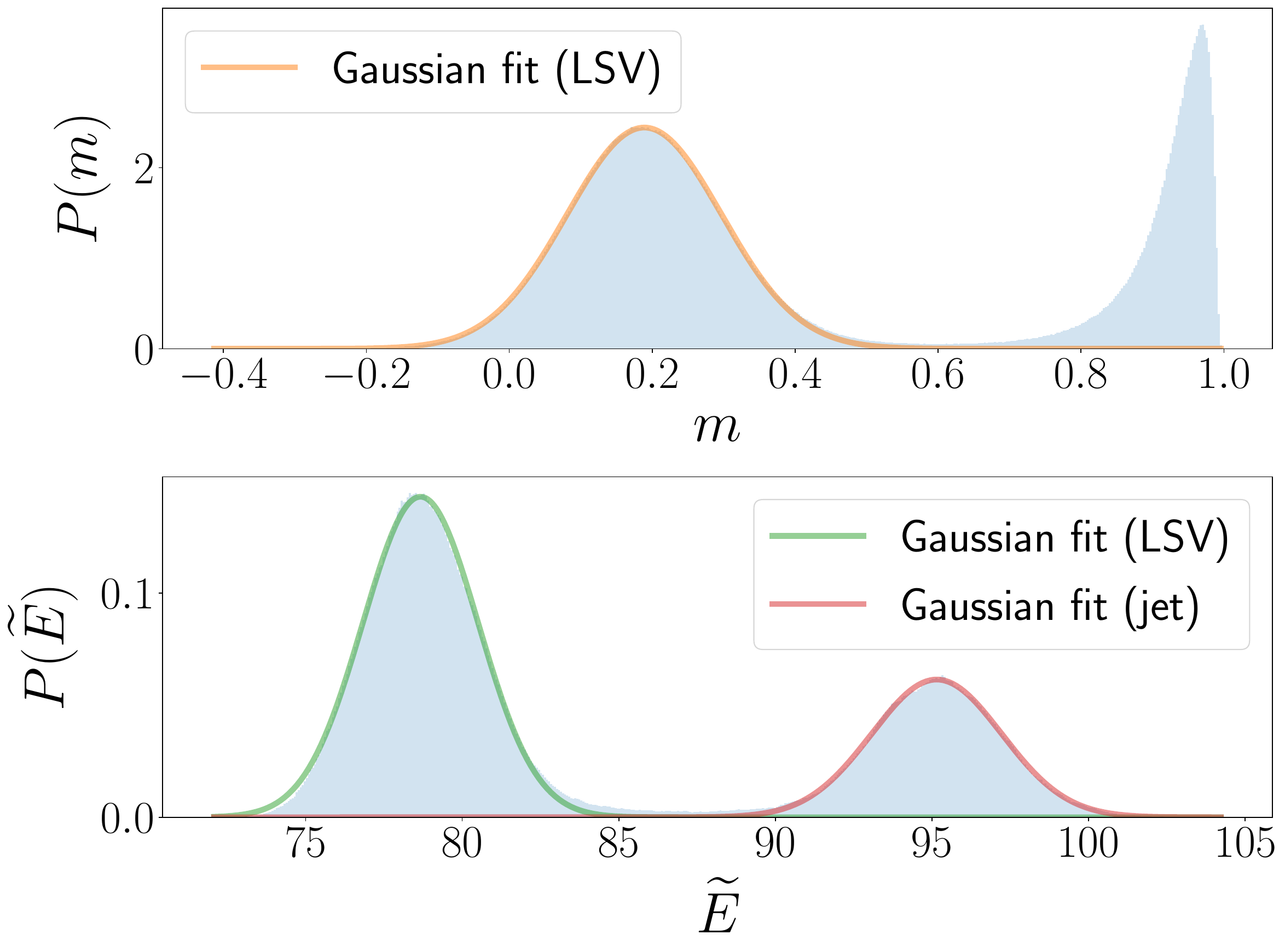}
\caption{Bimodal histograms of $m$ and $\widetilde{E}$ in a simulation with $\R=459$, $\tilde{\ell}=0.34$, and $\delta=1.065$ (set D1). The peak close to $m\approx1$ corresponds to the jet state, while the peak centered around $m\approx 0.2$ corresponds to the LSV state. 
In the kinetic energy histogram, the lower-energy peak corresponds to the vortex state, while the higher-energy peak corresponds to the jet state. It is interesting to note that the peaks in $P(m)$ and $P(\widetilde{E})$ are approximately Gaussian, except for the peak at $m\approx 1$, where $m$ is bounded above by $1$ by definition, leading to an asymmetric peak shape.}
\label{fig:scalar_distribution}
\end{figure}
\begin{figure}
    \centering
   (a) \hspace{7.5cm} (b) \includegraphics[width=\textwidth]{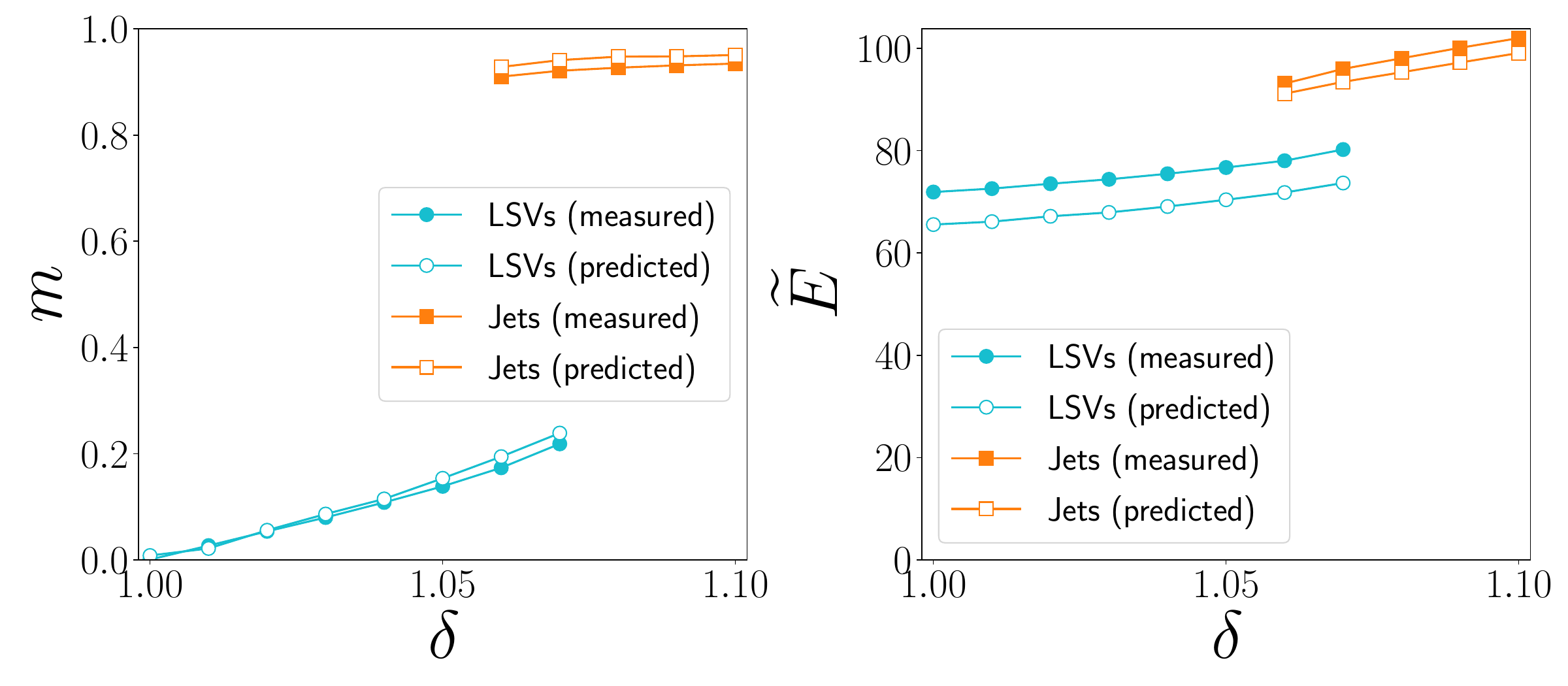}
    \caption{Comparison between the values of the polarity $m$ (panel (a)) and nondimensional energy $\widetilde{E}$ (panel (b)) at different 
 values of $\delta$ in \avkrev{set D1 at $\R=459$, $\tell=0.34$} measured in DNS and the predicted values from Eqs.~(\ref{eq:theo_pred_E}) and (\ref{eq:theo_pred_m}). (a)  Polarity $m$ in the LSV state is predicted correctly albeit with a slight overestimate. (b) In agreement with Fig.~\ref{fig:transfer_rates}(b), $\widetilde{E}$ in the jet state is close to that in the large-scale modes, while in the LSV state there is a $\lesssim 10\%$ difference between the large-scale energy and the total energy. Sufficient statistics in the jet state were only obtained at aspect ratios $\delta\geq 1.05$ and in the LSV state for aspect ratios $\delta \leq 1.07$.}
\label{fig:model_verification}
\end{figure}

\subsection{Measurements of polarity $m$ and energy $\widetilde{E}$ from DNS}
In all simulations, we record time series of polarity $m$ and nondimensional energy $\widetilde{E}$, and compile histograms of these random variables in the turbulent flow. Figure~\ref{fig:scalar_distribution} shows the distributions $P(m)$ and $P(\widetilde{E})$  at $\delta=1.065$, $\R=459$ (set D1). These are clearly bimodal, indicating bistability between the jet and LSV states. This bimodality is observed over a range of $\delta$ for given $\R$ and $\tilde{\ell}$. The structure of the observed peaks is close to Gaussian (as verified by the fits shown in Fig.~\ref{fig:scalar_distribution}), except for the peak associated with the jet state near $m=1$, which is asymmetric due to the upper bound $m\leq 1$. It is well known that near-Gaussian statistics are obeyed by many observables in 2D turbulence, cf. \cite{boffetta2000inverse,l2002quasi}.

Figure~\ref{fig:model_verification} shows a comparison between the time-averaged values of $m$ and $\widetilde{E}$ in the jet and LSV states, measured in DNS at $\R = 459$, and the corresponding predictions from Eqs.~(\ref{eq:theo_pred_E}) and (\ref{eq:theo_pred_m}).  In Fig.~\ref{fig:model_verification}(a) one sees a satisfactory agreement between the DNS and the predicted values of $m$ based on the large-scale energy balance. In panel (b), one further sees that the total kinetic energy in the jet state is close to the predicted value, while in the LSV state the observed kinetic energy is slightly higher than the contribution from the large-scale modes alone due to finite $\R$ effects, in agreement with Fig.~\ref{fig:transfer_rates}(b). Overall, we conclude that there is agreement between the DNS results and our predictions based on large-scale energy balance. 

 

\begin{figure}
\includegraphics[width = 0.75\textwidth]{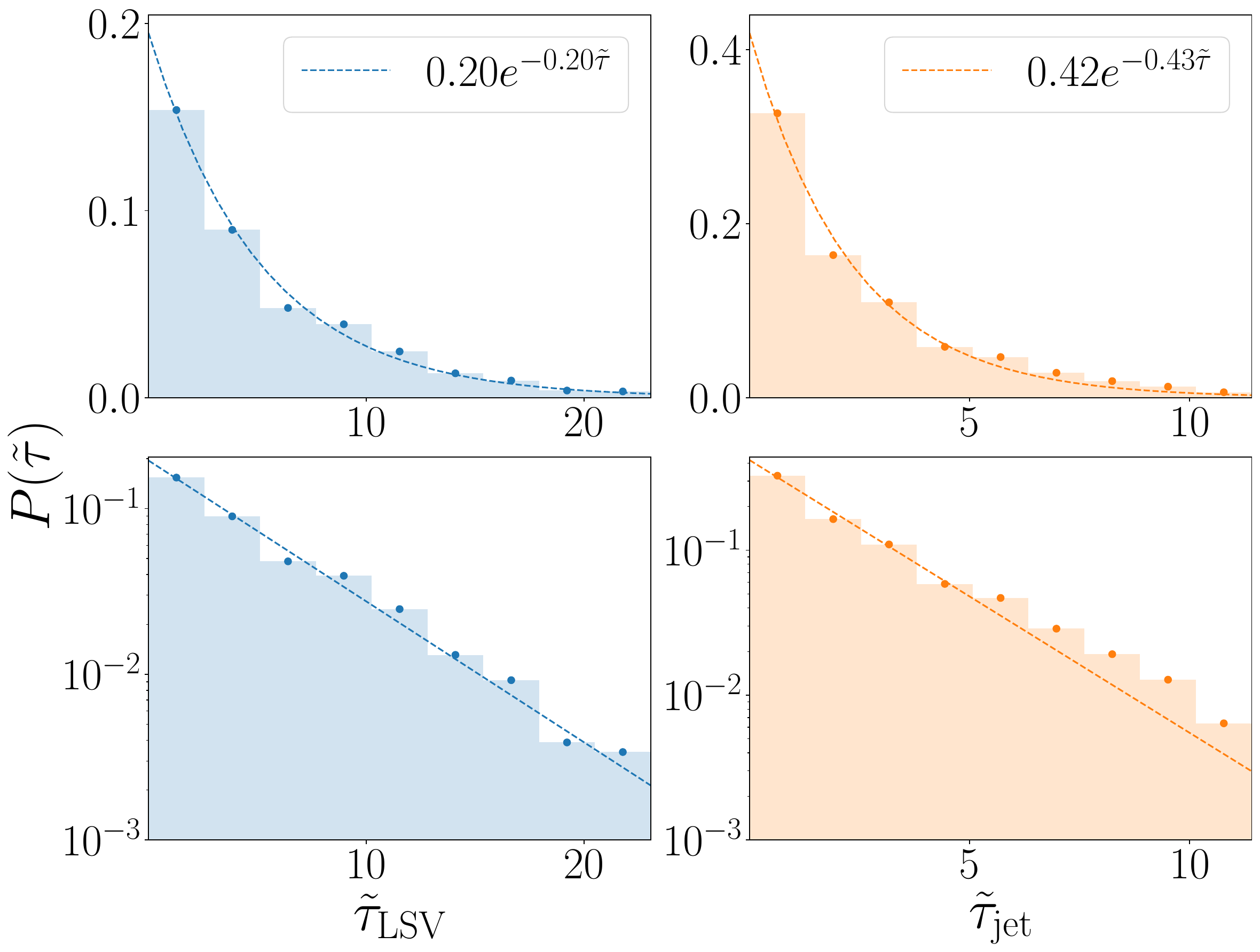}
\caption{\avkrev{Histogram of the lifetimes of LSVs and jets in a simulation with $\delta = 1.07$ and $\mathrm{Re} = 275$ (under-sampled tails are cut off). Data is from simulation set R. The dashed lines show an exponential fit which is in good agreement with the data. The second row shows the $y$ axis on a logarithmic scale to further validate the approximately exponential trend. The histograms shown are based on a total of $741$ realizations of jets and $805$ realizations of LSVs.}}
\label{fig:tau_distribution}
\end{figure}


\section{Lifetime statistics\label{sec:random}}

A key quantity to be measured in this problem is the nondimensional lifetime $\tilde\tau$ of the large-scale structures, i.e. the length of the time intervals between two consecutive transitions. Depending on the type of flow structure present, we denote the lifetime as $\tilde\tau_{\mathrm{LSV}}$ or $\tilde\tau_{\mathrm{jet}}$. We measure these times in units of the viscous time $t_\nu$ defined in Eq.~(\ref{eq:tnu}). For all lifetimes reported here, we subtract $0.5t_{\nu}$ to account for the duration of the transition, see also Fig.~\ref{fig:transitions_scalar}.  

Since the transitions take place at random times, $\tilde\tau$ is a random variable \avkrev{whose statistics we will analyze in the following.} 
The lifetimes of each structure are observed to approximately follow exponential distributions, as shown in Fig.~\ref{fig:tau_distribution}, indicating that they arise from a memoryless Poisson process. Similar findings have been reported for a wide variety of metastable states in turbulent flows, e.g. \cite{van2019rare,gome2020statistical,gome2022extreme,de2022bistability,de2022discontinuous,wang2023lifetimes}.

In the following, we examine the dependence of the mean lifetime $\langle\tilde\tau\rangle$ of LSVs and jets on parameters. Specifically, we focus on the impact of the nondimensional control parameters   $\delta$, $\R$ and $\tilde{\ell}$ on $\langle\tilde\tau_{\mathrm{LSV}}\rangle$ and $\langle\tilde\tau_{\mathrm{jet}}\rangle$.

\subsubsection{Aspect ratio \avkrev{$\delta=L_x/L_y$}}
As $\delta$ is increased from unity, the system transitions from preferring the LSV state to preferring the jet state. One expects that this is associated with $\langle\tilde\tau_{\mathrm{LSV}}\rangle$ decreasing as $\delta$ increases, and $\langle\tilde\tau_{\mathrm{jet}}\rangle$ increasing as $\delta$ increases. To examine this, we measured $\langle\tilde\tau\rangle$ in two sets of simulations (D1 and D2) with ${\rm Re}=459$ and ${\rm Re}=344$, respectively, varying $\delta$ from $1.045$ to $1.1$, as shown in Fig.~\ref{fig:tau_vs_delta}. We observe that $\langle\tilde\tau_{\mathrm{LSV}}\rangle$ and $\langle\tilde\tau_{\mathrm{jet}}\rangle$ intersect at $\delta = 1.07$ for $\R=344$ and at slightly lower $\delta$ for $\R=459$, which we can identify as the \textit{transition point} between the LSV and jet states. \avkrev{The data in Fig.~\ref{fig:tau_vs_delta} show that $\langle\tilde\tau_{\mathrm{jet}}\rangle$ increases by a constant factor as $\R$ increases, producing a simple shift on the logarithmic $y$ axis, while $\langle \tilde\tau_{\mathrm{LSV}}\rangle$ increases its slope as $\R$ increases (Fig.~\ref{fig:tau_vs_delta}). This appears to result in a slight shift in the transition point towards smaller $\delta$ as $\R$ increases, although this needs to be verified with additional simulations.}  

\begin{figure}[!htbp]
\includegraphics[width = 0.75\textwidth]{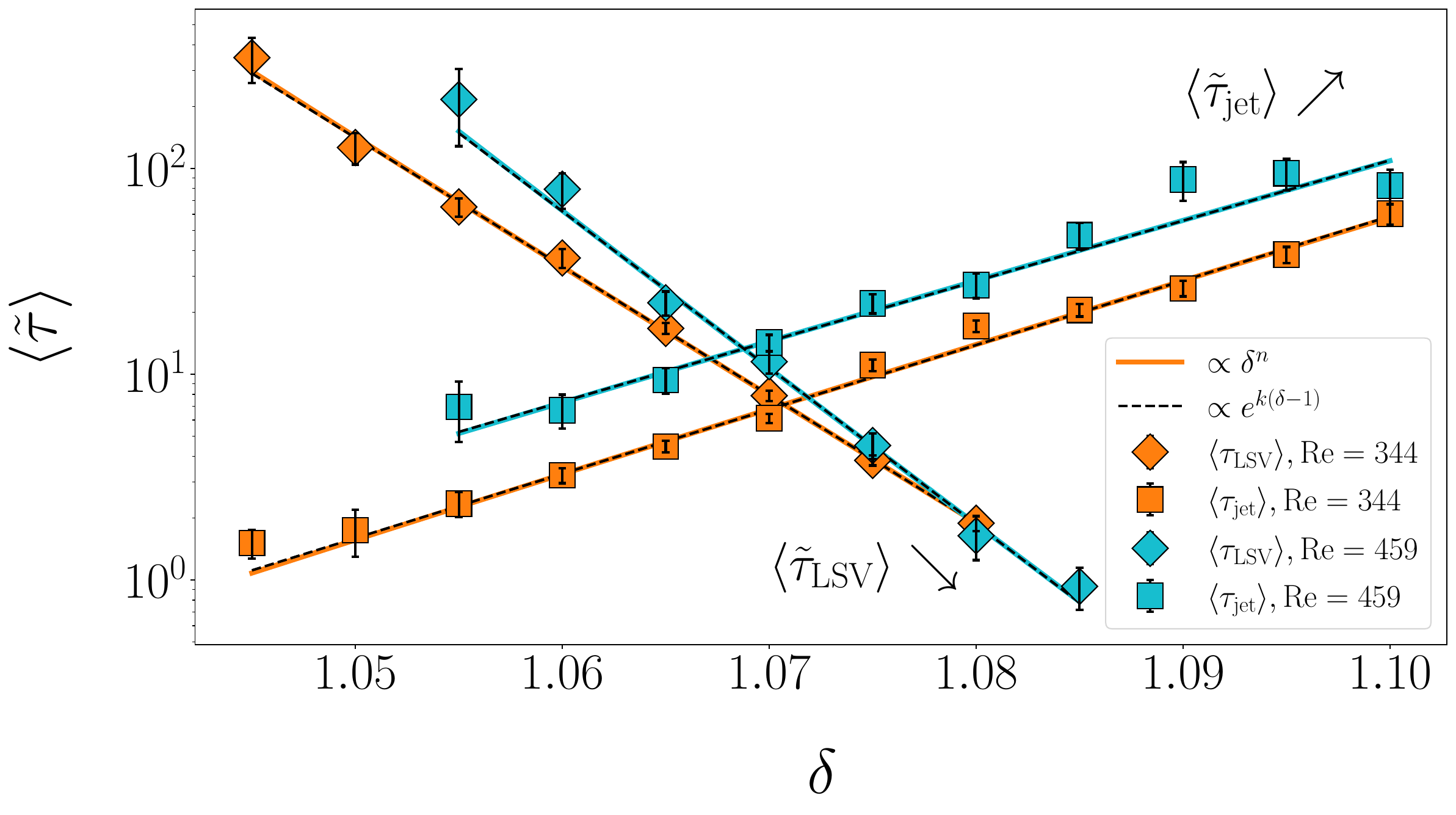}
\caption{Mean lifetimes of LSVs (circles) and jets (squares) versus $\delta$ from simulations at two distinct Reynolds numbers, $\R=344$ (D2, blue) and $\R=459$ (D1, orange). Error bars correspond to $\Delta \langle \tilde \tau\rangle = \Delta \tilde\tau /\sqrt{N}$, where $N$ is the sample size and $\Delta \tilde\tau$ is the sample standard deviation. {A minimum number of $20$ (at $\delta =1.05$) up to a maximum number of $500$ (at $\delta =1.07$) consecutive transitions were observed.} Dashed lines, which indicate best-fit exponential relations, are indistinguishable from the solid lines representing best-fit power-law relations. This is due to the small range of $\delta\approx 1$ over which the transitions are observed, implying $\delta \approx \log(\delta) + 1$.  }
\label{fig:tau_vs_delta}
\end{figure}

In Fig.~\ref{fig:tau_vs_delta}, we see that $\langle\tilde\tau_{\mathrm{LSV}}\rangle$ and $\langle\tilde\tau_{\mathrm{jet}}\rangle$ are consistent with the fitted exponential function of $\delta$ (solid lines). A similar exponential relationship (Arrhenius law) has been observed, for instance, in the context of rare transitions between different numbers of jets in $\beta$-plane turbulence \cite{bouchet2019rare}. However, we cannot discard the possibility that the dependence is a power law with a large exponent (dashed lines), due to the narrow range in $\delta$ over which bistability is observed. 
Both results, either power-law or Arrhenius form, differ from the case of rare transitions in anisotropic three-dimensional turbulence \cite{de2022bistability} (between 3D turbulence and LSVs), where the time scales of LSV formation or decay clearly show a faster-than-exponential dependence on the control parameter.

As discussed in \cite{de2022bistability}, another possibility is that the mean lifetime of a given structure diverges at some parameter threshold. This is numerically very difficult to ascertain since the required simulation times become extremely long, which has led to false conclusions in the past, cf. \cite{avila2010transient}. For instance, we cannot exclude, based on our data, that the mean lifetime of the jet state might diverge at some $\delta=\delta_c\geq 1$, where $\delta_c$ is an aspect ratio with which $\langle\widetilde\tau_\mathrm{LSV}\rangle = \langle\widetilde\tau_\mathrm{jet}\rangle$, for certain choices of $\R$ and $\tell$. However, in Sec.~\ref{sec:tell} we show that at least for the special case $\tell=1/\sqrt{5}$, the lifetime of the LSV state is finite and jets can form spontaneously even in the square domain at $\delta=1$.

\begin{figure}
\includegraphics[width = 0.75\textwidth]{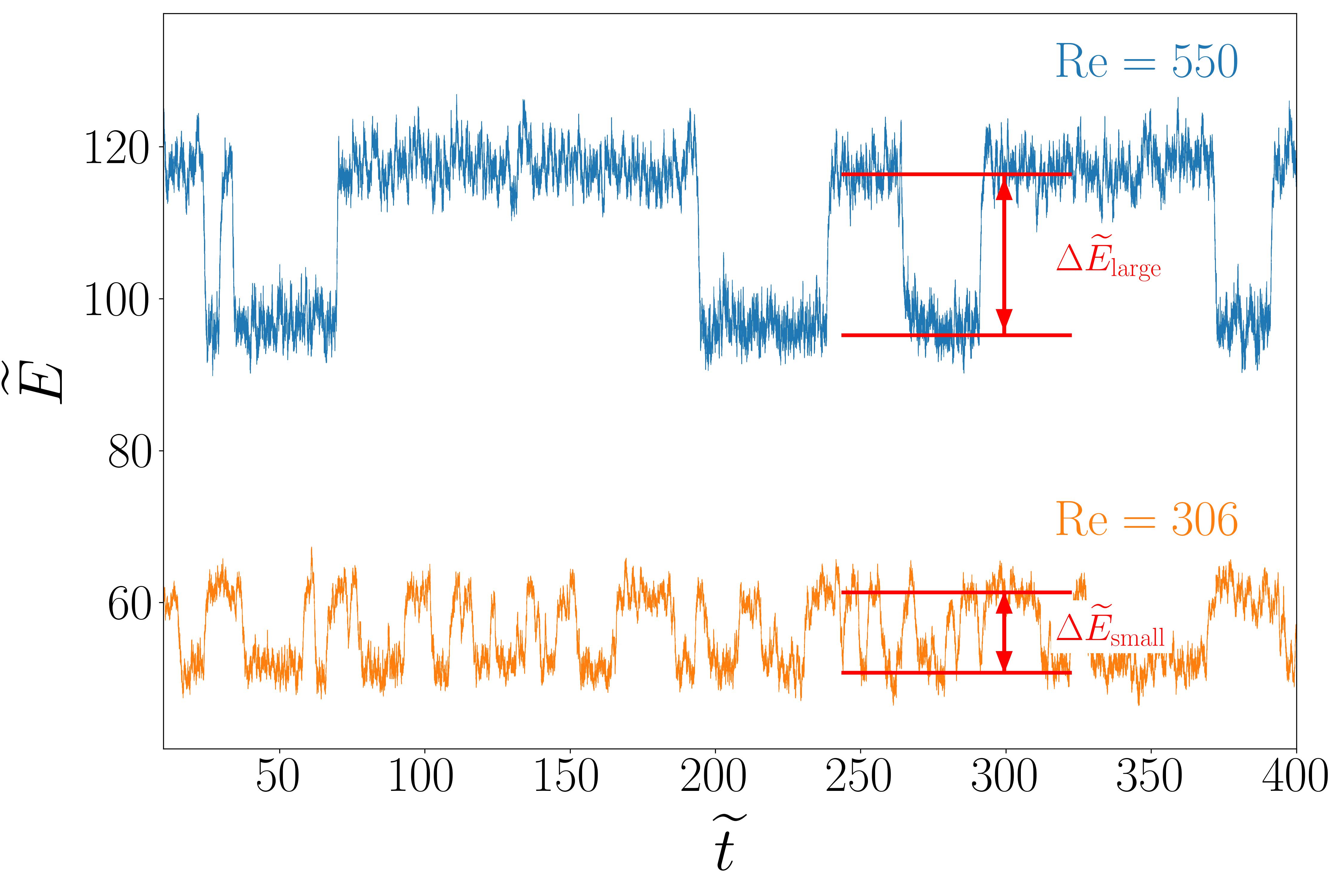}
\caption{Time series of nondimensional energy $\tilde{E}=E/E_f$ in simulations with different $\mathrm{Re}$ at $\delta = 1.07$ (set R). Smaller $\mathrm{Re}$ corresponds to lower energies and a smaller difference between $E_{\mathrm{LSV}}$ and $E_{\mathrm{jet}}$, leading to more frequent random transitions.}
\label{fig:Re}
\end{figure}

\begin{figure}
    \centering
    \includegraphics[width=0.8\textwidth]{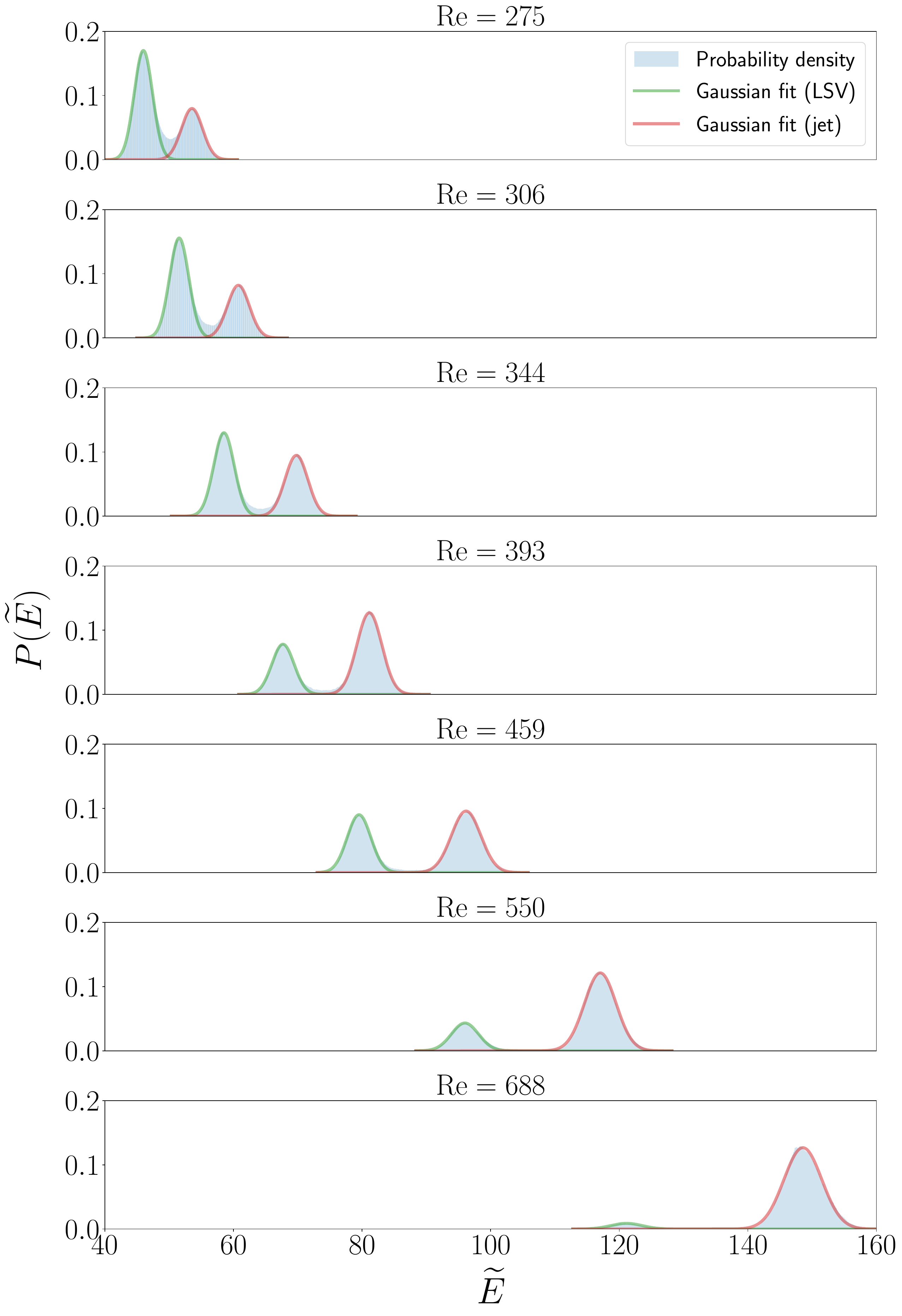}
    \caption{\avkrev{Histograms of total kinetic energy (nondimensionalized by the forcing scale energy) for runs in set $\mathrm{R}$ at $\delta=1.07$, $\tell=0.34$. The kinetic energy of jets and LSVs inceases with $\R$, and so does the amplitude of fluctuations around either state.}}
    \label{fig:histograms_vs_Re}
\end{figure}

\subsubsection{Reynolds number $\mathrm{Re}$}
\avkrev{The dependence of mean lifetimes on the Reynolds number $\R$ in this system is somewhat counter-intuitive. One might expect that larger values of $\R$ correspond to larger fluctuation amplitudes and thus more frequent transitions. However, as we will see below, transition rates decrease (i.e. lifetimes increase) as $\R$ is increased as a consequence of an increasing energy gap between the two structures.}

In our setting, $\mathrm{Re}$ is closely related to the total kinetic energy $E$, with larger $\mathrm{Re}$ corresponding to a higher $E$, cf. Eq.~(\ref{eq:theo_pred_E}). This is consistent with the time series of kinetic energy at different Reynolds numbers shown in Fig.~\ref{fig:Re}, which show that energy of both jets and LSVs increases with $\R$, as does the energy gap. \avkrev{Figure~\ref{fig:histograms_vs_Re} shows histograms of nondimensional energy at different $\R$ from simulation set R, revealing that the energy of the jet and LSV states, as well as the gap between them [Fig.~\ref{fig:Egap_vs_Re}(a)], all increase with $\R$, as does the amplitude of the fluctuations in the total energy (dominated by the large-scale contribution), although the energy gap increases more rapidly than the amplitude of the total energy fluctuations [Fig.~\ref{fig:Egap_vs_Re}(b)].} 
Quantitatively, at a fixed $\delta$ and sufficiently large $\R$, we expect the gap between the LSV energy $E_{\mathrm{LSV}}$ and the jet energy $E_{\mathrm{jet}}$ to be approximately proportional to $\R$ based on the predictions of Sec.~\ref{sec:energy_balance_theory}. This expectation is confirmed in Fig.~\ref{fig:Egap_vs_Re}(a). A smaller energy gap compared to the forcing energy scale causes the system to transition more frequently between the two structures because the system needs to dissipate/absorb less energy when the transition happens. Therefore, lower $\mathrm{Re}$ result in more frequent transitions and shorter $\tilde\tau$ than higher $\R$, as observed in Fig.~\ref{fig:Re}.


\begin{figure}
    \centering
       (a) \hspace{7.5cm} (b) 
    \includegraphics[width=0.49\textwidth]{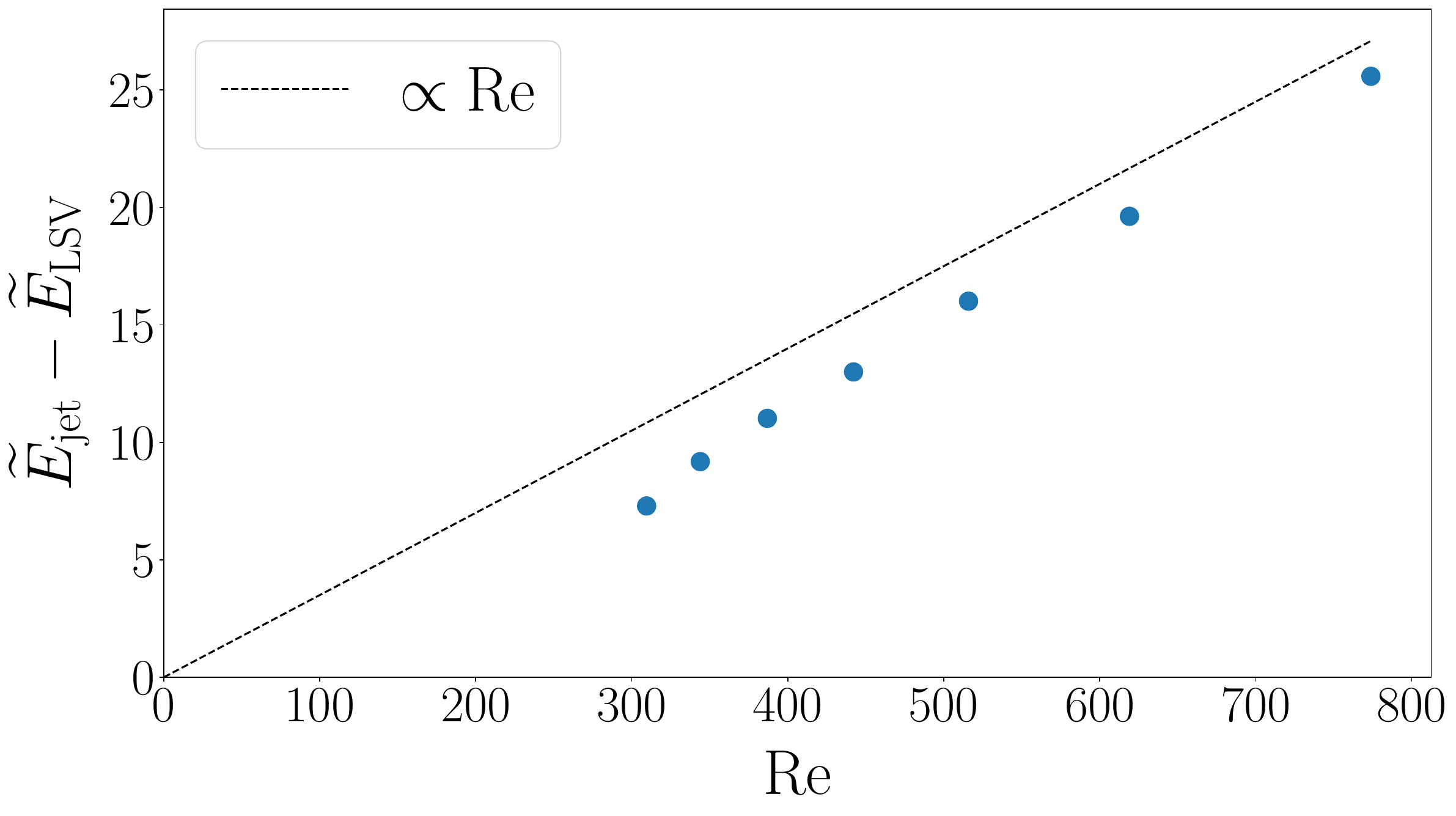}
        \includegraphics[width=0.49\textwidth,height=0.275\textwidth]{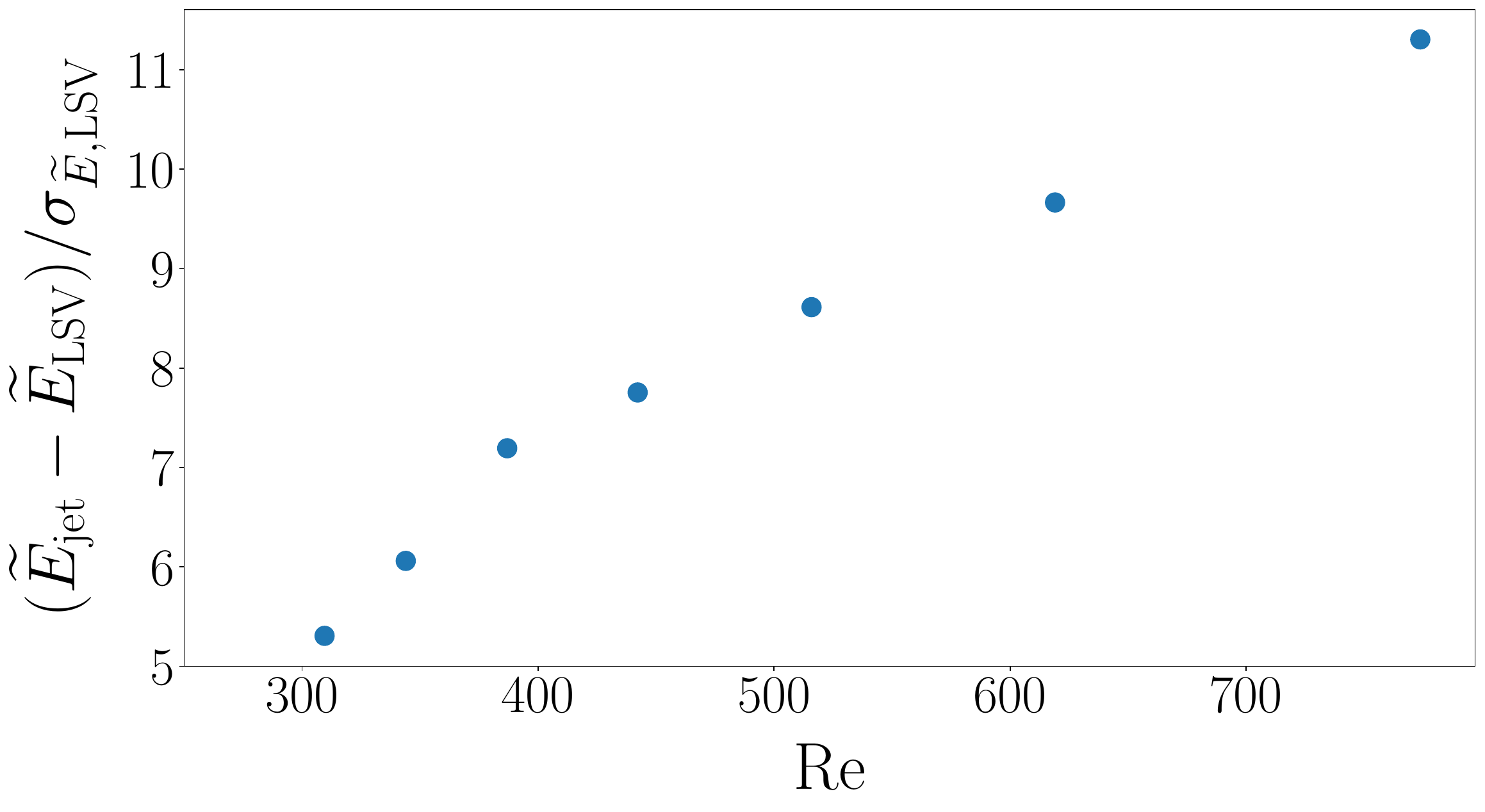}
    \caption{\avkrev{(a) Energy gap between LSV and jet states versus $\R$ at $\delta = 1.07$, $\tell=0.34$ (set R). A close to linear trend is observed at large $\R$. (b) The energy gap divided by the standard deviation of the total energy in the LSV state (cf. Fig.~\ref{fig:histograms_vs_Re}) increases with $\R$.}}
    \label{fig:Egap_vs_Re}
\end{figure}

\avkrev{Next, we quantified the dependence of $\langle\tilde\tau\rangle$ on the Reynolds number $\R$ based on the runs in set R, varying $\R$ from $275$ to $688$.} Figure~\ref{fig:tau_vs_RE} shows a lin-log plot of the resulting mean lifetimes of LSVs and jets versus $\R$. \avkrev{One observes that there is again an approximately exponential dependence, as shown by the best fits for LSVs and jets (solid lines). 
\avkrev{While the exponential fit appears to be consistent with our data, a power-law dependence also provides an acceptable, albeit slightly less accurate fit (not shown), giving $\langle \widetilde{\tau}_{\mathrm{jet}}\rangle \propto \R^{3.65}$ and $\langle \widetilde{\tau}_{\mathrm{LSV}}\rangle\propto \R^{0.98}$ at $\delta=1.07$ and $\tell=0.34$.} Incidentally, a similar power-law dependence on the Rayleigh number $\mathrm{Ra}$ has recently been reported in \cite{wang2023lifetimes} for metastable states in windy convection, with $\langle \tau\rangle \propto \mathrm{Ra}^{4.05}$.}

\begin{figure}[!htbp]
\includegraphics[width = 0.75\textwidth]{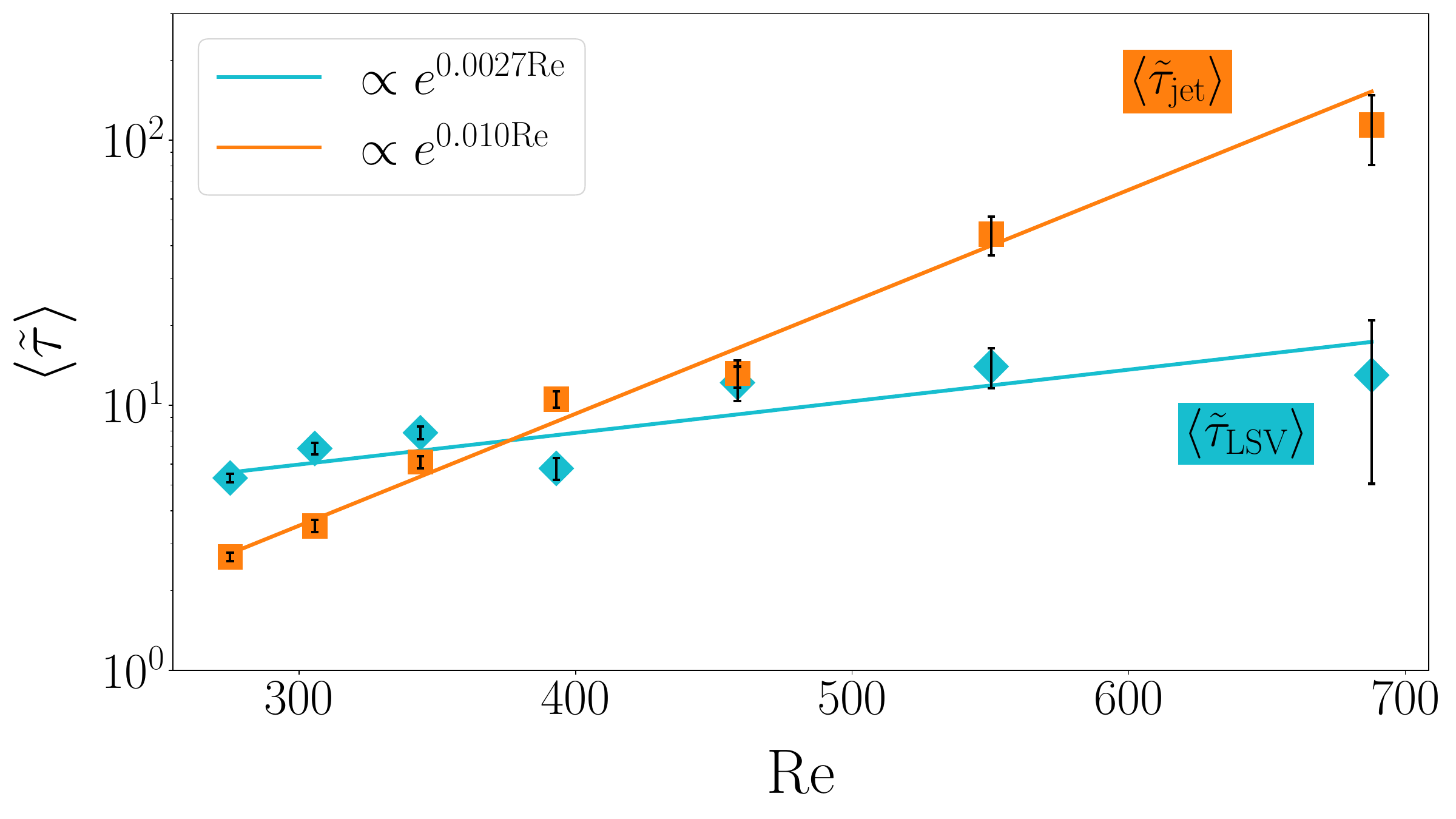}
\caption{Lin-log plot of the mean lifetimes of LSVs and jets versus $\R$ \avkrev{from simulation set R ($\delta=1.07$ and $\tell=0.34$). Solid lines 
show exponential best fits, which are consistent with the lifetime data for jets and LSVs, respectively.  A minimum number of $50$ (at $\R=688 $) up to a maximum number of $1000$ (at $\R =275$) consecutive transitions were observed.} }
\label{fig:tau_vs_RE}
\end{figure}
\begin{figure}
\includegraphics[width = 0.75\textwidth]{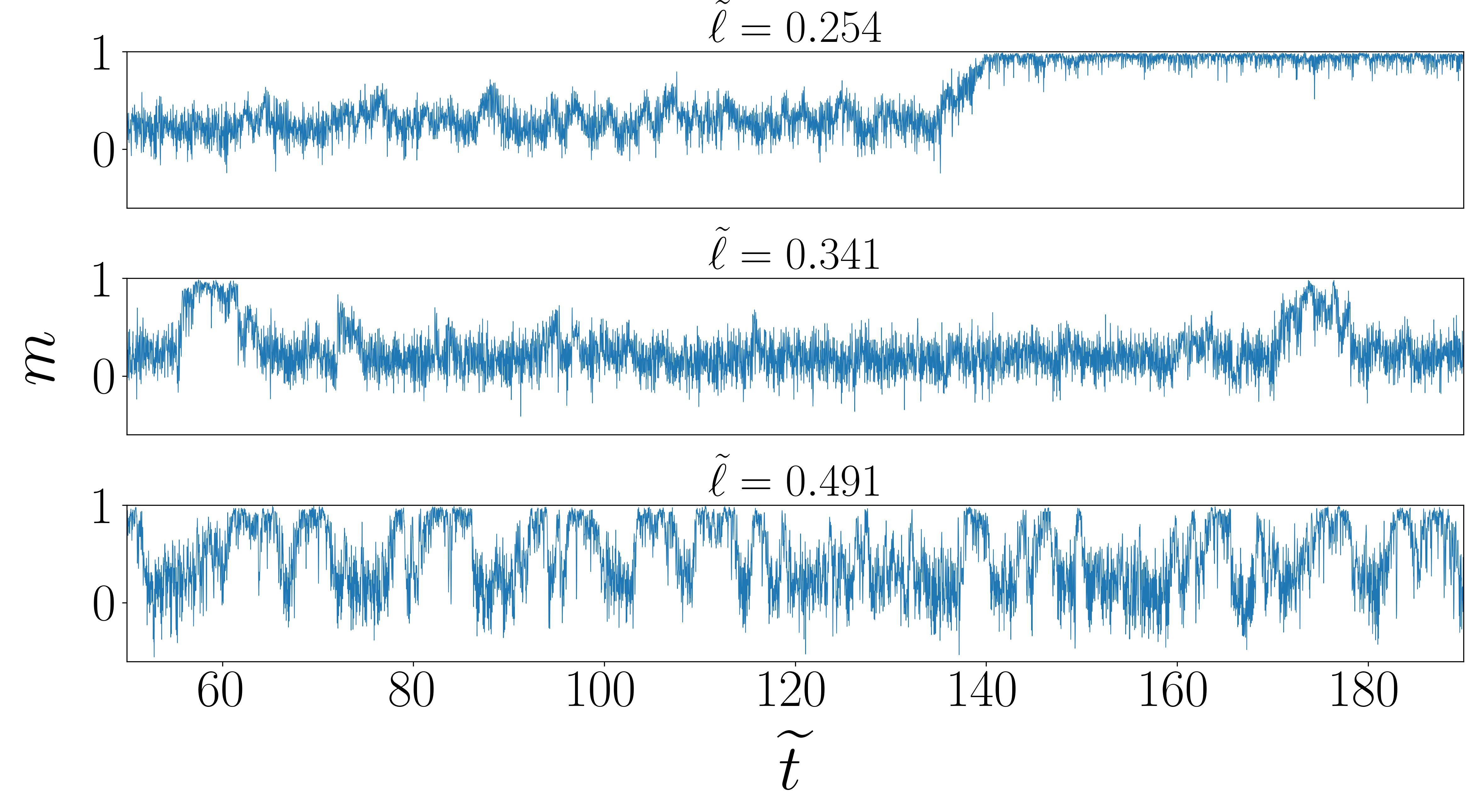}
\caption{Segments of time series of the polarity $m$ in simulations from set $\mathrm{L}$ with different nondimensional forcing scales \avkrev{$\tilde{\ell}=\ell/L_y$} (increasing top to bottom) for $\delta=1.06$ and $\R=69$. Larger forcing scales lead to substantially more frequent random transitions. }
\label{fig:impact_scale_separation_on_transition_times}
\end{figure}

\begin{figure}
    \centering
\includegraphics[width=0.75\textwidth]{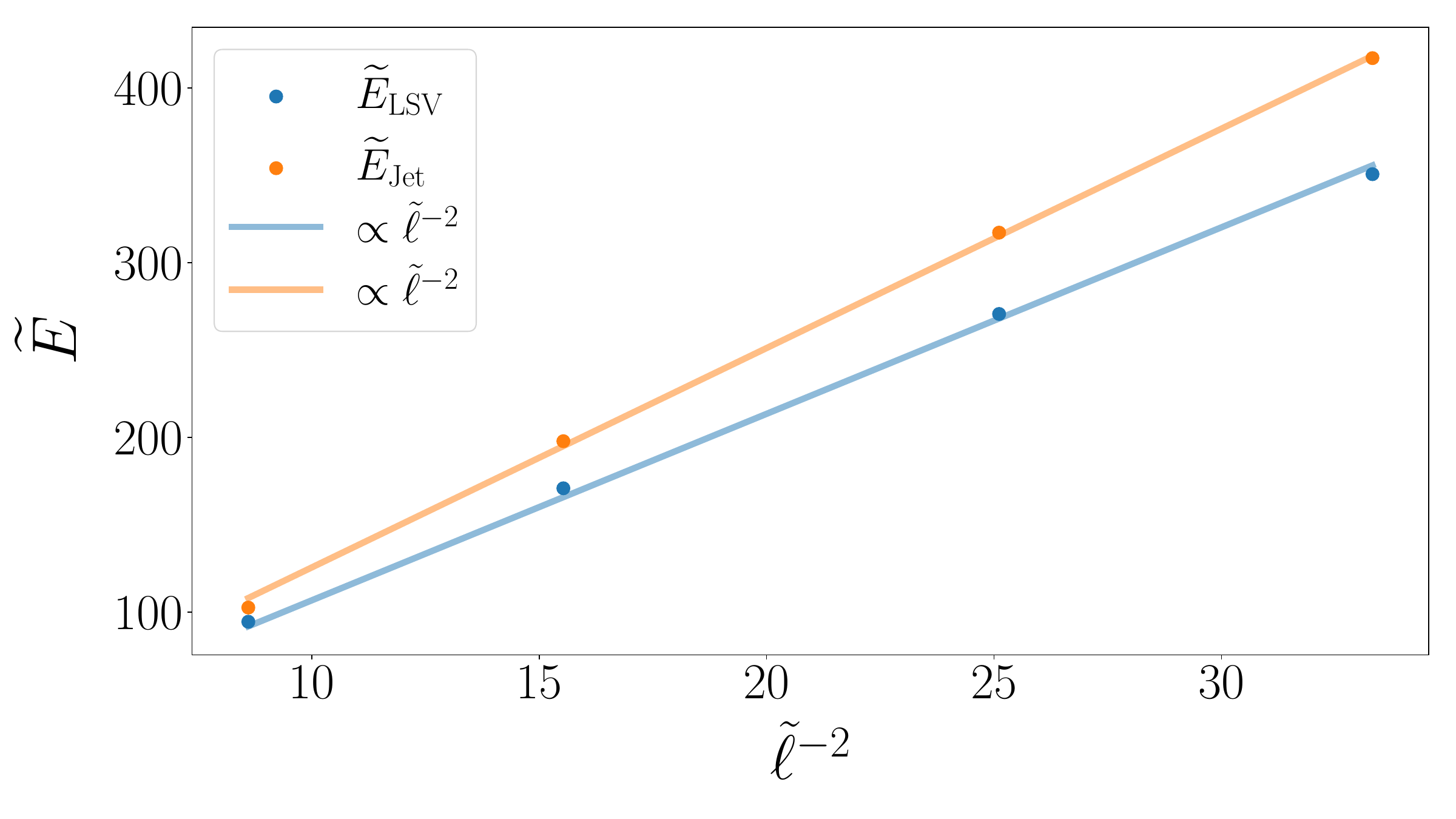}
    \caption{\avkrev{Nondimensional energy $\widetilde{E}_\mathrm{jet}$ of jet state and nondimensional energy $\widetilde{E}_\mathrm{LSV}$ of LSVs versus $\tell$ from simulation set $\mathrm{L}$ ($\R=69$, $\delta=1.06$). Both quantities scale approximately as $\tell^{-2}$, in agreement with large-scale energy balance as expressed in Eq. (\ref{eq:theo_pred_E_nondim}). Thus the energy gap $\widetilde{E}_\mathrm{jet}-\widetilde{E}_\mathrm{LSV}$ grows rapidly as $\tell$ decreases.} }
    \label{fig:energies_lsv_jets_vs_tell}
\end{figure}
\begin{figure}
\includegraphics[width = 0.75\textwidth]{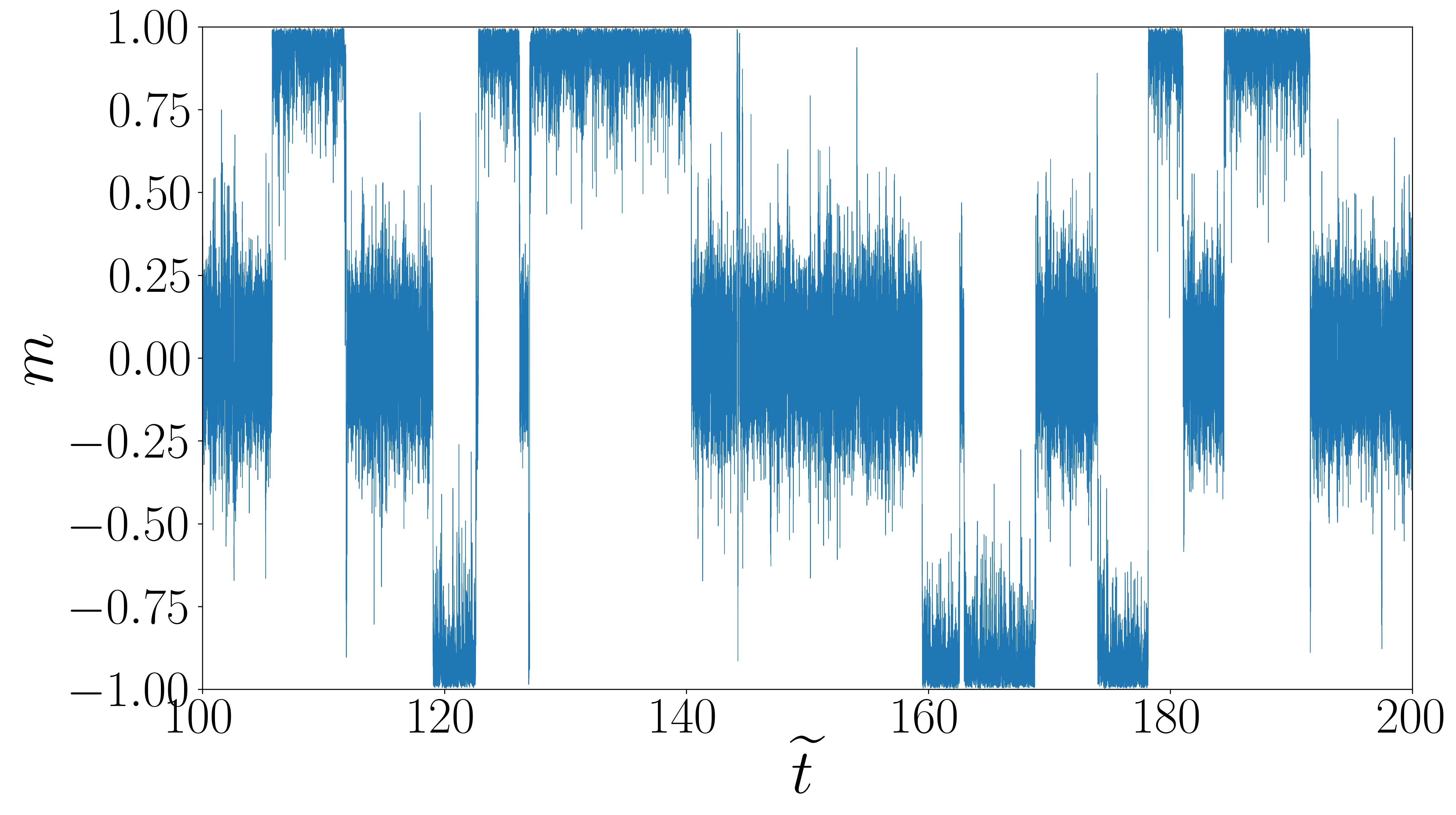}
\caption{Time series of the polarity $m$ on a square domain (aspect ratio $\delta=1$) with nondimensional forcing scale $\tilde{\ell} = 1/\sqrt{5}$ and Reynolds number $\mathrm{Re} = 550$ (from simulation set T). The polarity $m$ randomly jumps between $-1$, $0$, and $+1$, indicating the coexistence of three types of large-scale structures in the domain. The jets in both directions are present in the system for approximately the same fraction of the simulation time, as expected from the symmetry under $x\leftrightarrow y$.}
\label{fig:tristability}
\end{figure}

\subsubsection{Nondimensional forcing scale $\avkrev{\tilde{\ell}=\ell/L_y}$\label{sec:tell}}
Finally, the nondimensional  ratio $\tilde{\ell} = \ell/L_y$ between the forcing and domain scales plays a critical role in the transitions. Figure~\ref{fig:impact_scale_separation_on_transition_times} shows that the results are highly sensitive to \avkrev{$\tilde{\ell}$, based on runs from simulation set $\mathrm{L}$}: the rate at which transitions occur between LSVs and jets decreases sharply with $\tell$: at $\tell=0.491$, there  are many transitions in the time interval shown ($\sim 120 t_\nu$), while the system switches between jets and LSVs four times at $\tell=0.341$ and only once at $\tell=0.254$ in the same time interval. Here, we mention that in the work of Bouchet and Simonnet \cite{bouchet2009random}, the forcing was concentrated near wave number $k=2$, i.e. $\ell=\pi$, with $L_x=2\pi \delta$ and $L_y=2\pi$ \cite{private_comm_Bouchet_Simonnet}, corresponding to $\tell\approx 0.5$ in our notation. A sensitive dependence of transition times on parameters was previously reported in \cite{bouchet2014non}. 

\avkrev{Figure \ref{fig:energies_lsv_jets_vs_tell} shows that both jet and LSV energies are approximately proportional to $\tell^{-2}$, in agreement with large-scale energy balance [cf. Eq.~\eqref{eq:theo_pred_E_nondim}]. This implies that $\widetilde{E}_\mathrm{jet}-\widetilde{E}_\mathrm{LSV}$, namely the size of the energy gap  between the two states relative to the forcing energy, also increases approximately as $\tell^{-2}$ when $\tell$ is reduced. This observation suggests that, similar to 
the decrease of transition rates with increasing Reynolds number, the decrease in the mean transition rate with decreasing $\tell$ appears to be related to the sharp increase in the energy gap. However, the highly sensitive dependence of lifetimes on $\tell$ prevents the type of detailed, quantitative analysis of the dependence of the mean lifetimes on $\tell$ that was done for their Reynolds number and aspect ratio dependence.  }

\avkrev{In summary, we find that mean lifetimes depend approximately exponentially on the domain aspect ratio $\delta$ (i.e. the anisotropy parameter), while for a given aspect ratio and $\tell$, the mean lifetimes also depend approximately exponentially on $\R$ (which is approximately proportional to the energy gap measured in units of the forcing energy), while the mean lifetimes also sharply increase when the energy gap widens due to a decreased nondimensional forcing scale $\tell$ (while keeping $\R$ and $\delta$ fixed). These observations suggest that an underlying Arrhenius law would be consistent with the measured exponential dependence of mean lifetimes on parameters although we are unaware of any theoretical arguments supporting this result. Should a quantity analogous to a quasi-potential exist for the present problem, our simulations indicate it must necessarily depend on $\delta$ (which controls the relative mean lifetimes of the two structures) and on the energy gap (controlling the absolute mean lifetimes of both structures). The forcing energy $E_f$ is likely an important measure of the fluctuation amplitude, although the fluctuations of the total energy (dominated by large scales) may also be relevant and may need to be taken into account.}

Next, we briefly discuss a special case that arises when the forcing is concentrated near $\tell = 1/\sqrt{5}$. In this case, shown in Fig.~\ref{fig:tristability}, a square domain with $\delta=1$ exhibits {\it tristability} and spontaneous transitions are observed between the LSV state and jet states in either the $x$ or $y$ direction. This contrasts starkly with the usual condensate state found in square domains, where typically a highly stable pair of counter-rotating LSVs is observed \cite{boffetta2012two}. We stress that this tristable behavior in a square domain was only seen when the forcing range included the scale $L_y/(2\pi\sqrt{5})$. We have explicitly verified that the metastable jets in the square domain do not form when forcing at $\tell\geq 1/\sqrt{5}$; specifically, we tested $\tell=1/2,1/\sqrt{2},1$ and also showed that this phenomenon does not occur for forcing scales smaller than $\tell=1/\sqrt{5}$. While the precise origin of this phenomenon is as yet unknown, we note that the wave number triads
\begin{equation}
    \mathbf{k}_{1,1} +  \mathbf{k}_{1,0} + \mathbf{k}_{-2,-1} = 0, \qquad \mathbf{k}_{1,1} + \mathbf{k}_{0,1} + \mathbf{k}_{-1,-2} = 0,
\end{equation}
which couple either one of the large-scale modes to a forcing-scale mode, facilitate a nonlinear interaction between the $\mathbf{k}_{1,0}$ and $\mathbf{k}_{0,1}$ modes and the forcing scales via the mediator mode $\mathbf{k}_{1,1}$, which is the closest scale to the gravest modes, permitting a strong spectrally local interaction. A full quantitative explanation of the observed dynamics remains to be elucidated. {It is interesting to note that similar turbulent jets alternating between the $x$ and $y$ directions in a square domain were also observed in rotating convection \cite{de2022discontinuous}.} 

\section{Transition Trajectories\label{sec:transitions}}
In addition to the lifetime statistics discussed above, it is also of interest to examine the transitions themselves. In particular, we investigate whether the transition from LSVs to jets and vice versa are time-reversible. In general, transition trajectories between distinct attractors in multi-stable systems are asymmetric. For instance, in \cite{simonnet2021multistability}, the transitions between states with 2 and 3 jets in $\beta$-plane turbulence follow different paths in Fourier space in different directions ($2\to3$ jets or $3\to 2$ jets). This motivates us to examine the similarity between transitions in the two directions in our system.
\begin{figure}
\includegraphics[width = 0.9\textwidth]{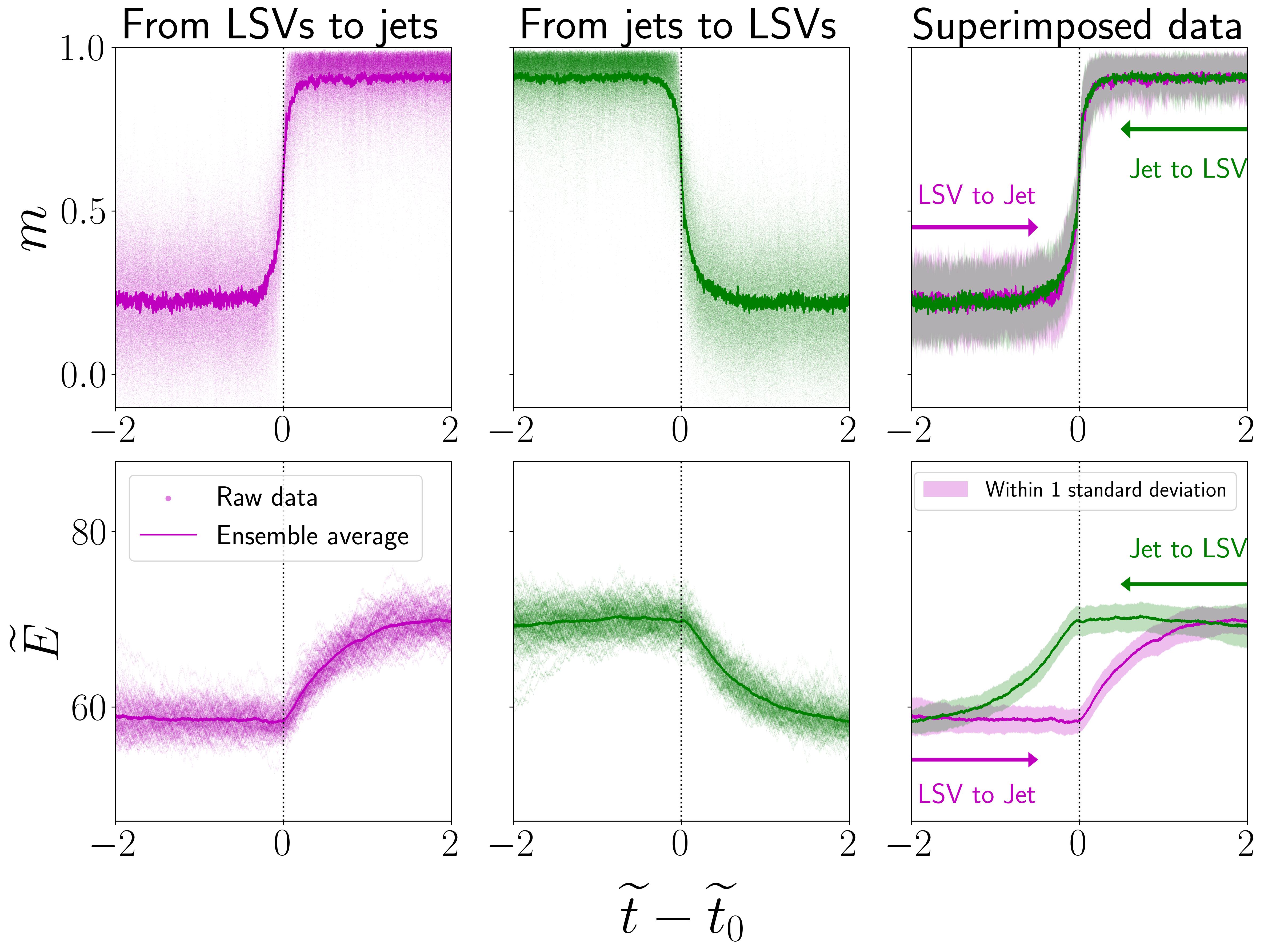}
\caption{Top row: Evolution of the polarity $m$ defined in Eq.~(\ref{eq:polarisation}) during transitions from LSVs to jets and vice versa, shifted in time such that the $m$ values coincide at nondimensional times $\widetilde t=\widetilde{t}_0$, where $m(\,\widetilde{t}_0\,)$ corresponds to the minimum of $P(m)$. The rightmost panel shows the overlap of the first two, where the time axis in the middle panel was reflected about $\widetilde t=\widetilde{t}_0$. The scatter points show the instantaneous system states in different realizations, while the solid lines represent the ensemble average. Bottom row: same for the nondimensional kinetic energy $\widetilde E$. Vertical dashed lines highlight $\widetilde t=\widetilde{t}_0$. Data are taken from a simulation in set D2 with $\delta = 1.07$, $\tell=0.34$ and $\mathrm{Re} = 344$, including 121 instances of transitions from LSVs to jets and 129 from jets to LSVs.}
\label{fig:transitions_scalar}
\end{figure}

\subsubsection{Evolution of $m$ and $\widetilde{E}$ during transitions}
As established in Sec.~\ref{sec:structures}, the polarity $m$ and the kinetic energy $\widetilde{E}$ serve as key indicators of the system state. Here, we focus specifically on the temporal evolution of these quantities during the transitions from jets to LSVs and vice versa.

We consider a long-time simulation in the bistable range, with a large number of transitions from LSVs to jets and vice versa. Figure~\ref{fig:transitions_scalar} shows superimposed sections of the time series of $m$ and $\widetilde{E}$ centered on the point during transitions, at nondimensional time $\widetilde{t}=\widetilde{t}_0$, where $m$ minimizes $P(m)$, cf. Fig.~\ref{fig:scalar_distribution}. Hence, by construction, all the time series shown in Fig.~\ref{fig:transitions_scalar} coincide at nondimensional time $\widetilde{t}=\widetilde{t}_0$. Point clouds indicate the instantaneous states of the system during individual transitions, while the solid lines indicate an average over the ensemble of all observed transitions (the data was additionally smoothed before the ensemble average was computed).

The top row of Fig.~\ref{fig:transitions_scalar} shows that the polarity rapidly transitions from LSVs to jets and vice versa. Superimposing the two ensemble-averaged trajectories by reflecting the time axis in the middle panel about $\widetilde{t}=\widetilde{t}_0$, one obtains the right panel, which shows that $m$ approximately follows the same transition path in either direction, within one standard deviation. By contrast, the lower panel shows that the kinetic energy $\widetilde{E}$ displays a drastically different transition behavior. Kinetic energy is approximately constant up to the shifted time $\widetilde{t}=\widetilde{t}_0$ (defined in terms of $m$ as described above). Then, the energy increases or decreases on a viscous time scale to adapt to the equilibrium value appropriate for the newly formed structure, given by Eqs.~(\ref{eq:energy_balance_steady_state}). When both curves are again superimposed by reflecting the middle panel about $\widetilde{t}=\widetilde{t}_0$ (right panel), it becomes apparent that the time delay in the energy evolution results in a markedly asymmetric, i.e. irreversible, transition behavior.

The distinct evolution of $m$ and $\widetilde{E}$ reveals the two-stage nature of the fluctuation-induced transitions in this system. First, a rapid nonlinear transfer of energy takes place, transforming one large-scale structure into the other, as indicated by $m$, and the kinetic energy $\widetilde{E}$ subsequently adjusts to the new equilibrium value. \avk{The role of the nonlinear transfer during transitions can also be analyzed directly by measuring the energy transfer rate $T_{a,b}$ defined in Eq.~(\ref{eq:Tab}). Figure \ref{fig:time_series_Tab} shows ensemble-averaged time series of $T_{1,0}$ and $T_{0,1}$ (nondimensionalized by the energy injection rate $\epsilon$) during transitions from LSVs to jets and vice versa. Away from the transition, the transfer rate is constant, corresponding to the fractions $\alpha_{1,0}$ and $\alpha_{0,1}$ of energy reaching the large-scale modes, cf. Eq.~(\ref{eq:transfer_rate_averaged}). In contrast, during the transition, an overshoot or an undershoot is seen in the time series, reflecting the rapid change in the large-scale energies $E_{1,0}$ and $E_{0,1}$.}

\begin{figure}
    \centering
    \includegraphics[width=0.9\textwidth]{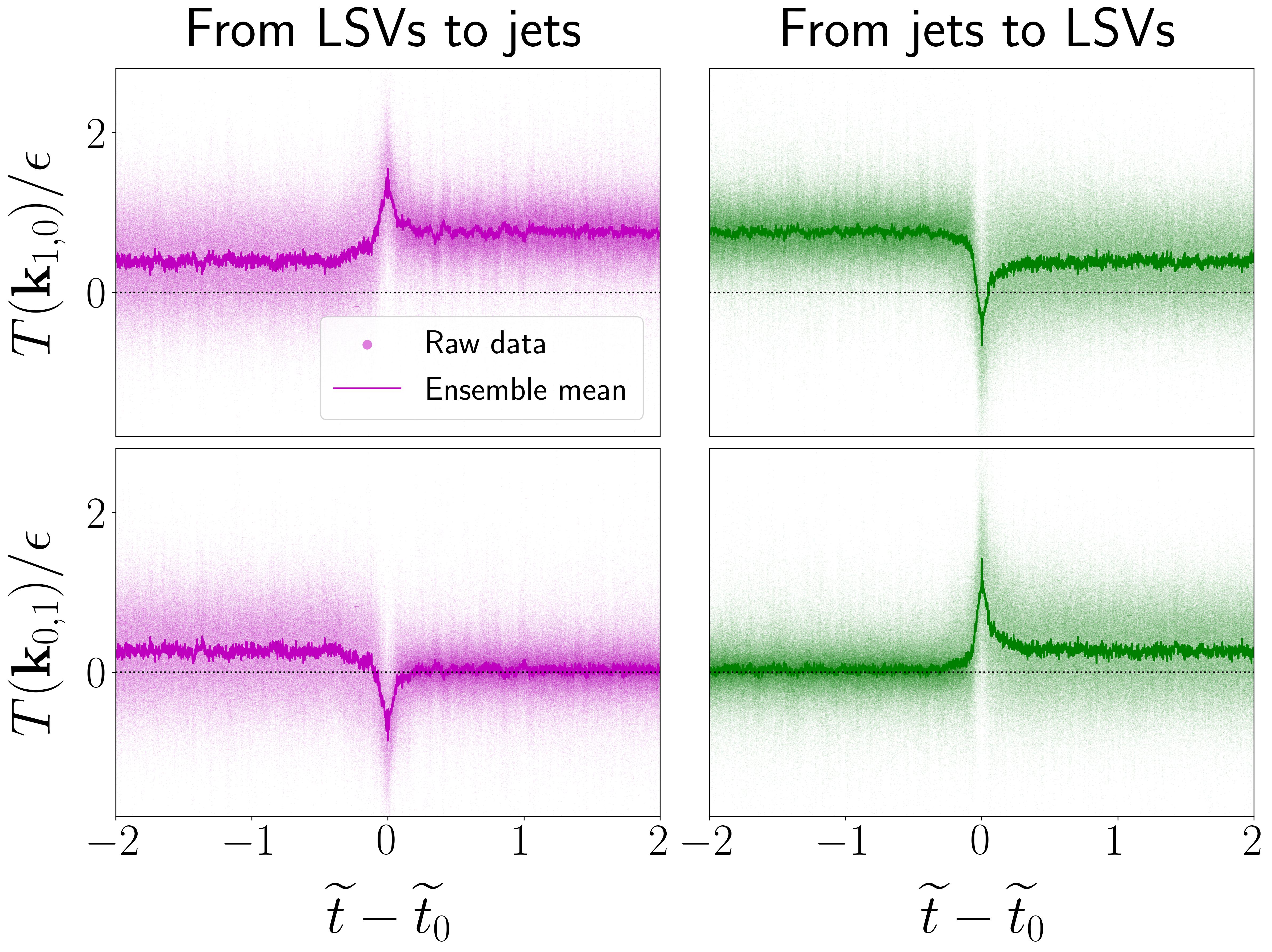}
    \caption{Time series of the nonlinear transfer $T_{1,0}$ (top row) and $T_{0,1}$ (bottom row), nondimensionalized by the injection rate $\epsilon$, during transitions shown over the same time axis as in Fig.~\ref{fig:transitions_scalar}. 
    Scatter points stand for the raw data (smoothed in time for clarity) and solid lines indicate the ensemble-average over a large number of transitions. The constant value of the ensemble averaged $T_{a,b}$ away from the transition corresponds to $\alpha_{a,b}$ in Eq.~(\ref{eq:transfer_rate_averaged}), while a pronounced overshoot/undershoot is seen during the transition, indicating the presence of large nonlinear transfer rates that lead to the rapid redistribution of energy reflected in the evolution of $m$ during transitions shown in Figs.~\ref{fig:transitions_scalar} and \ref{fig:m_E_phase_space_portrait}. Data are taken from the same simulation as in Fig.~\ref{fig:transitions_scalar}. }
    \label{fig:time_series_Tab}
\end{figure}

\begin{figure}
    \centering
\includegraphics[width=0.7\textwidth]{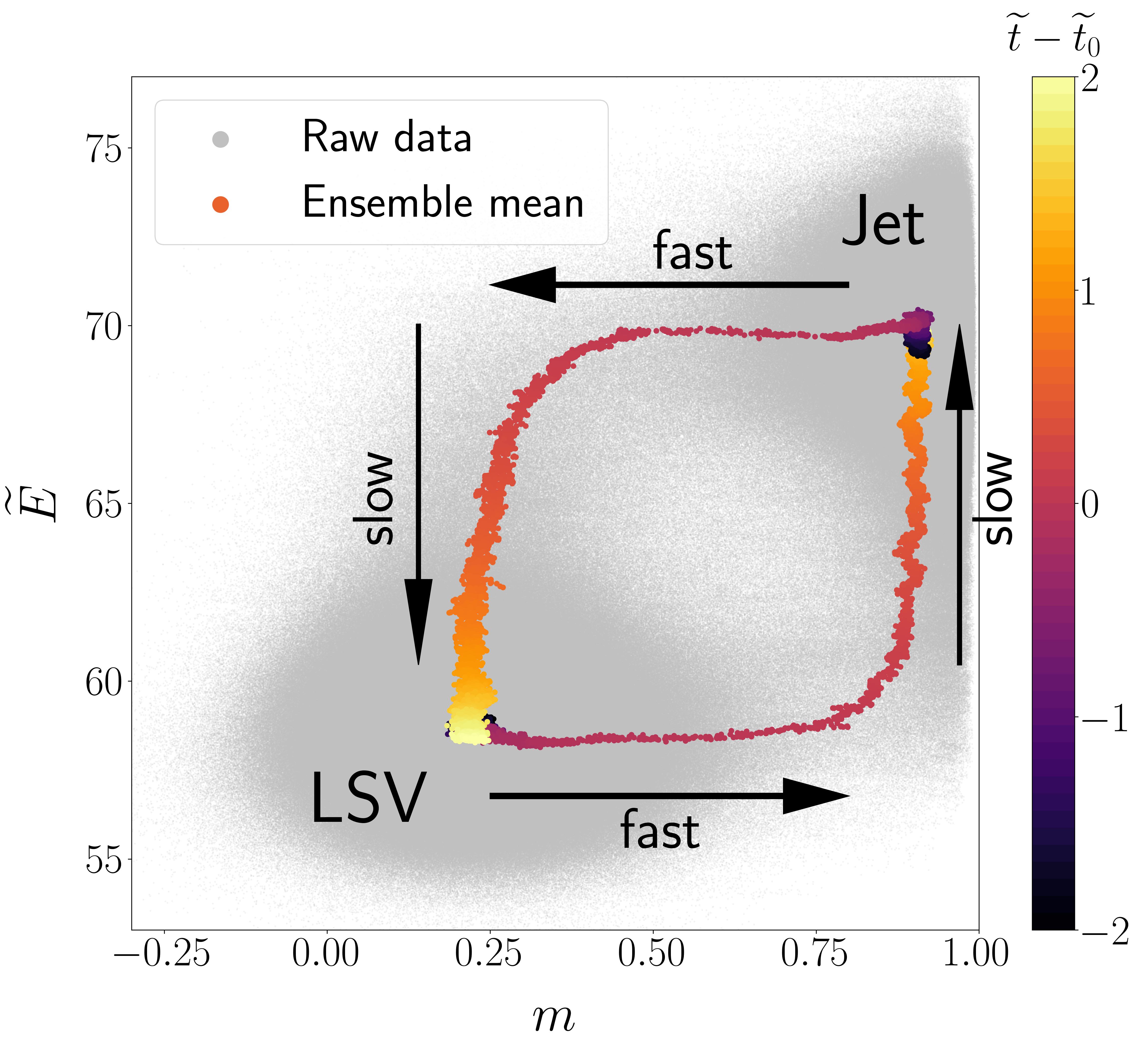}
    \caption{Dynamics in phase space $(m,\widetilde{E})$ within the bistable range from a run in set D2 ($\delta = 1.07$, $\R = 344, \tell=0.34 $, same simulation as in Fig.~\ref{fig:transitions_scalar}). Raw data are represented as grey point clouds, while ensemble-averaged transition trajectories are shown by colored, solid curves. Colors indicate the nondimensional time elapsed during transitions (corresponding to the abscissa of Fig.~\ref{fig:transitions_scalar}). Labels mark LSV and jet states, with arrows representing the counterclockwise transition direction consisting of a fast phase (small change in $\tilde{t}$) and a slow phase (larger change in $\tilde{t}$). The phase portrait is qualitatively reminiscent of relaxation oscillations \cite{van1926lxxxviii}, but the period at which the system completes the loop is the sum of the lifetime of the jet and the lifetime of the LSV. }
    \label{fig:m_E_phase_space_portrait}
\end{figure}

Figure~\ref{fig:m_E_phase_space_portrait} shows a reduced phase space portrait of the bistable dynamics for a run in set D2 at $\R=344$, $\delta =1.07$ in the reduced phase space spanned by $(m,\widetilde{E})$ as two key global system properties. 
The grey point cloud indicates instantaneous system states, while the colored lines represent the ensemble-averaged transition trajectories from LSV to jet and vice versa. It can be seen that the system spends most of its time in the regions corresponding to LSVs and jets, separated by occasional transitions from one region to the other, as expected. Colors indicate the time elapsed during the transition relative to $\tilde{t}=\tilde{t}_0$. The evolution of the system from LSVs to jets and back to LSVs corresponds to a loop in this phase space that is always traversed counterclockwise, in a manner reminiscent of relaxation oscillations \cite{van1926lxxxviii}, with rapid changes in $m$ and slow changes in energy, but with the important difference that the cycle is completed within a random time period given by the sum of the lifetimes of jet and vortex states. Figure~\ref{fig:m_E_phase_space_portrait} highlights that transitions are initiated by a rapid change in the polarization $m$, followed by a slow relaxation in the kinetic energy $\widetilde{E}$ to the new state.
\begin{figure}
\hspace{-0cm} (a) \hspace{8cm} (b)\\
\includegraphics[width = 0.98\textwidth]{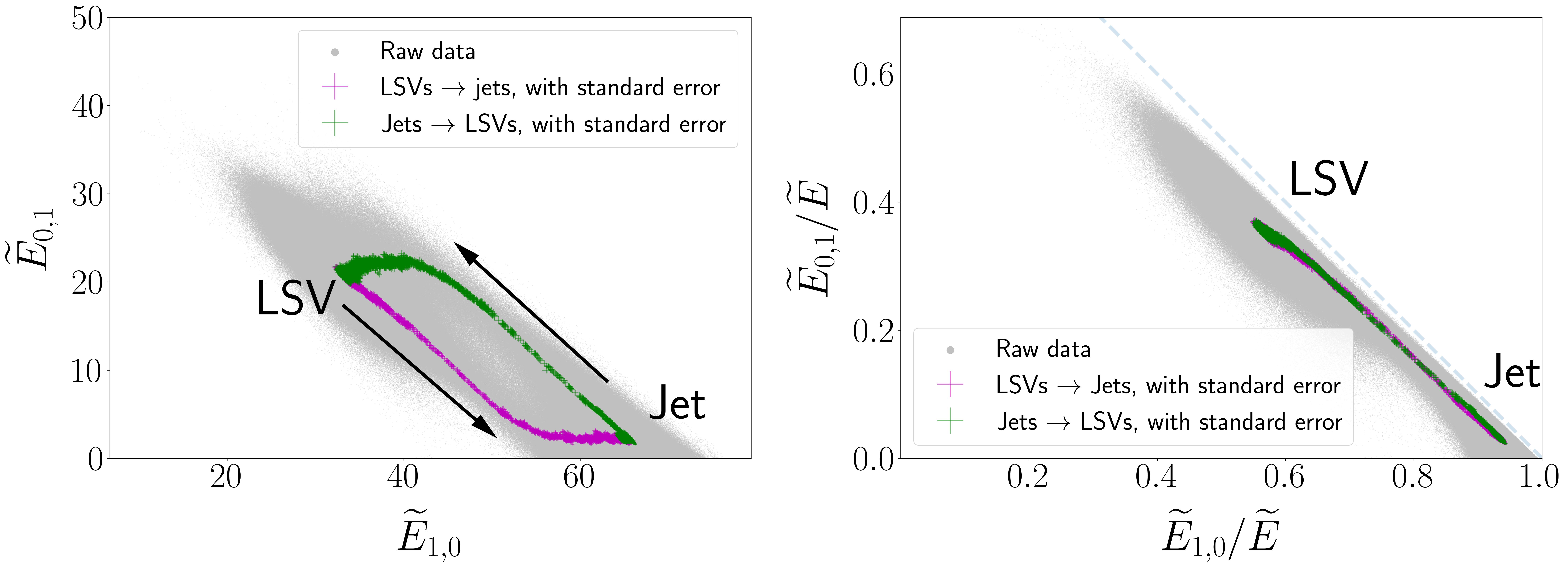}
\caption{Alternative phase space plots of the bistable system in Fourier space from a simulation in set D2 with $\delta = 1.07$ and $\R = 344$. (a) Absolute contributions to the energy $(\widetilde{E}_{1,0},\widetilde{E}_{0,1})$, whose sum approximates the energy $\widetilde{E}$. Grey points in both plots indicate the instantaneous system states, while the purple and green symbols indicate the averaged trajectories the system traverses during a transition. (b) Fraction of the energy contained in the large scales $(\widetilde{E}_{1,0}/\widetilde{E},\widetilde{E}_{0,1}/\widetilde{E})$, whose difference approximates the polarity $m$. The LSV and jet states have been highlighted by labels while the transition direction is indicated by arrows in panel (a). The dashed diagonal line in panel (b) is where $\widetilde{E}_{1,0}+\widetilde{E}_{0,1}=\widetilde{E}$.}
\label{fig:Fourier_density}
\end{figure}

\subsubsection{Evolution of energy in large-scale Fourier modes\label{sec:Fourier}}

In addition to $m$ and $\widetilde{E}$ defined in terms of the physical-space velocity field, the set of (complex) Fourier coefficients spans Fourier space and thus provides a full characterization of the fluid flow. To visualize the spectral energy distribution in the large-scale flow structures, we use the nondimensional energy in the modes $\pm \mathbf{k}_{a,b}$, denoted by $\widetilde{E}_{a,b}$, as defined in Eq.~(\ref{eq:Eab}).
We reiterate that, since the inverse energy cascade concentrates the energy in the largest-scale motions, the system is well described by the first few Fourier modes with the smallest wave numbers. Therefore the modal energies $\widetilde{E}_{1,0}$ and $\widetilde{E}_{0,1}$ are sufficient to characterize the state of the system. In addition, we also consider the \textit{fraction} of the total kinetic energy  $\widetilde{E}$ contained in $\widetilde{E}_{1,0}$ and $\widetilde{E}_{0,1}$ at any given time (recall that $\widetilde{E}$ changes between vortex and jet states), given by $\widetilde{E}_{1,0}/\widetilde{E}$ and $\widetilde{E}_{0,1}/\widetilde{E}$, respectively. 


 Figure~\ref{fig:Fourier_density} shows the system evolution in terms of $(\widetilde{E}_{1,0},\widetilde{E}_{0,1})$ (panel (a)) and $(\widetilde{E}_{1,0}/\widetilde{E},\widetilde{E}_{0,1}/\widetilde{E})$ (panel (b)), with the grey point cloud again indicating the set of all instantaneous system states and the purple/green lines corresponding to the ensemble averaged transition trajectories from LSV to jet and vice versa. It should be noted that the system spends most of its time in the regions corresponding to LSVs and jets, while occasionally traversing from one region to the other.

In Fig.~\ref{fig:Fourier_density}(a), a loop can be discerned in terms of the ensemble averaged transition trajectories, similar to that in Fig.~\ref{fig:m_E_phase_space_portrait}.
By contrast, in Fig.~\ref{fig:Fourier_density}(b), this loop collapses close to a dashed line indicating states where the entire energy is contained within the large-scale modes, i.e., $\widetilde{E}_{1,0}+\widetilde{E}_{0,1}=\widetilde{E}$. It can be seen that the jet states are somewhat closer to the dashed line than the LSV states, indicating that the remaining small-scale motions in the LSV state are more vigorous compared to the jet state. While this difference is small compared to the total energy, it does indicate a measurable difference between the energy spectra of the LSV and jet states. This is consistent with the spectra shown in Fig.~\ref{fig:spectra} and with the results shown in Fig.~\ref{fig:transfer_rates}.

\section{Transition Between Different Numbers of Jets}
\label{sec:jets}

\begin{figure}
\includegraphics[width=0.8\textwidth]{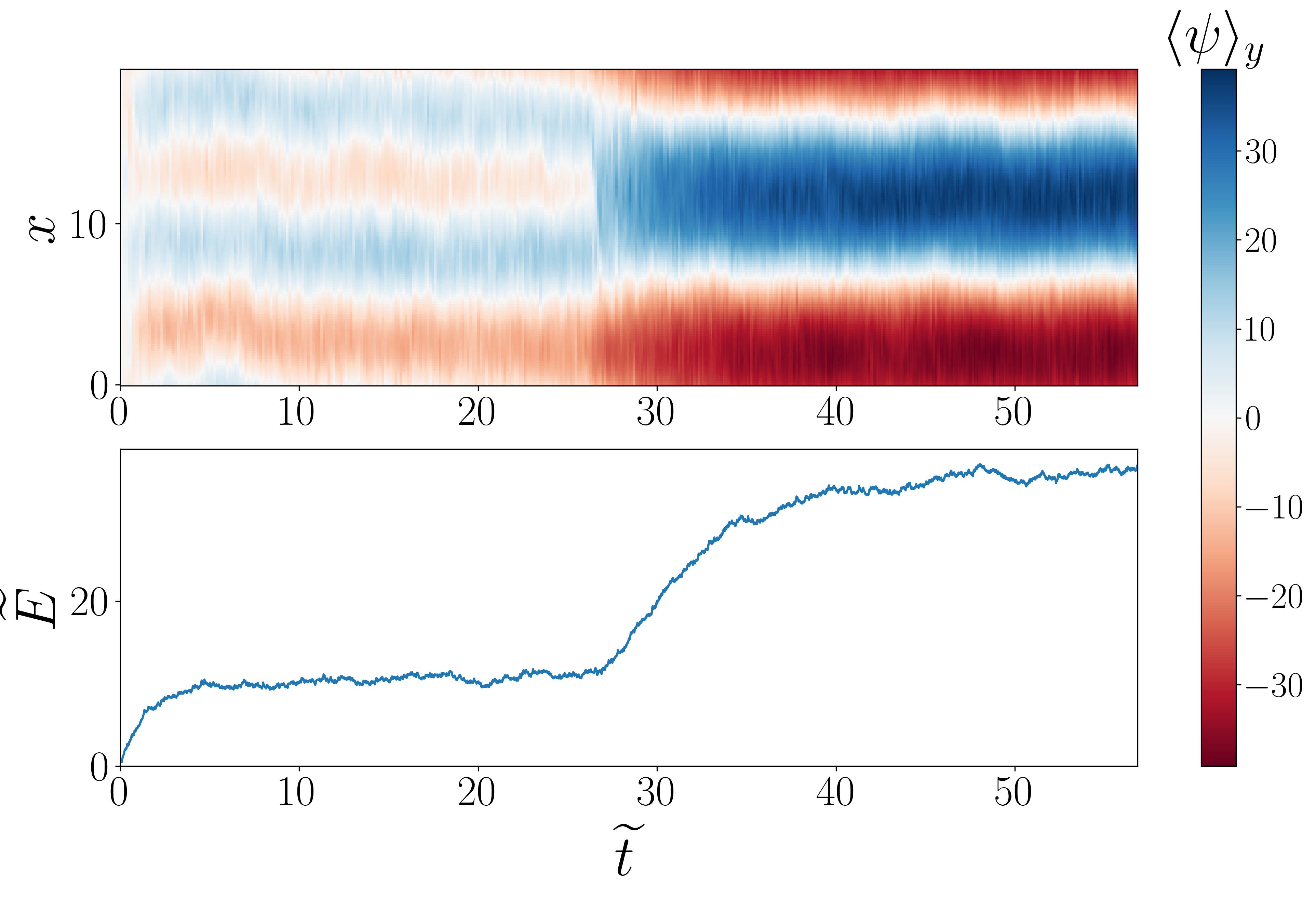}
\caption{(a) \avkrev{Space-time plot of $y$-averaged stream function from run $\mathrm{J}$ in a domain with aspect ratio $\delta \approx 3.1125$, at low Reynolds number $\R=55$ and $\tell =0.53$, showing atransition from two pairs of jets to a single pair of jets.} The flow amplitude of the jets in this transition increases greatly because the energy in the system is stored in larger-scale motions and is therefore less susceptible to viscous dissipation. (b) Kinetic energy time series associated with the transition reveals a significant energy gap overcome in the transition.}
\label{fig:2jets_4jets}
\end{figure}
When $\delta$ is increased further beyond the bistable interval discussed above, the number of jets emerging from the inverse cascade increases. Figure~\ref{fig:2jets_4jets}(a) shows a space-time plot of the stream function averaged in the $y$ direction $\langle \psi\rangle_y$, with the $x$ coordinate on the vertical axis and time in viscous units on the horizontal axis, from a run at $\delta\approx 3.1$, $\tell=0.53$, low Reynolds number $\R=55$, with a spatial resolution of $128\times 64$ grid points. A spontaneous transition from a state with two pairs of jets to a state with a single pair of jets is initiated around $\widetilde{t}\lesssim 30 $. Since the length scale doubles during the transition from two pairs of jets to a single pair of jets, the flow amplitude is significantly increased owing to reduced viscous dissipation, an effect seen in Fig.~\ref{fig:2jets_4jets}(b) as a drastic (close to fourfold) increase in kinetic energy during the transition. \avk{A similar coarsening-type evolution from two pairs of jets to a single pair of jets was observed in rapidly rotating convection in an anisotropic domain \cite{guervilly2017jets}.} 

{We also underline that a single pair of jets at $\delta >1$ corresponds to a build-up of energy at the largest available scale, also known as the gravest mode, which is in line with the expectation of standard theory \cite{smith1993bose}. However, when several pairs of jets are present in the system, the inverse cascade is arrested at a scale comparable to (\avkrev{but somewhat larger than}) the short side length ($L_y$ for $\delta>1$). The same phenomenology was observed in turbulent rotating convection \cite{julien2018impact,guervilly2017jets}, but there is so far no detailed explanation of this observed arrest of the inverse cascade at scales comparable to  the short side length}. {\color{black}Moreover, jets aligned with the short side of the domain also form in fixed-flux Rayleigh-B\'enard convection in a doubly periodic domain, where the secondary instability of an elevator mode generating a horizontal jet requires a small enough horizontal extent \citep{liu2023fixed}. }

In the presence of a dominant scale-independent damping mechanism, such as Rayleigh damping, the energy gap between states with different numbers of jets is removed, see e.g.~\cite{bouchet2019rare}. Previous studies of fluctuation-induced transitions between different numbers of jets include barotropic flows in the presence of the $\beta$ effect (latitudinal variation of the Coriolis force) but similar studies of anisotropy-generated jets remain to be carried out.


\section{Conclusion}
\label{sec:conclusions}
In this investigation, we made significant contributions to our understanding of fluctuation-induced transitions in anisotropic 2D turbulence by performing extensive DNS and acquiring statistical data on the impact of key system parameters on the dynamics. We have formulated and verified predictions for the energy and flow polarity of LSVs and jets based on large-scale energy balance. Furthermore, we have shown that the lifetimes of the large-scale structures are exponentially distributed, indicating that they derive from a memoryless process, as observed for various metastable states in other turbulent flows \cite{van2019rare,gome2020statistical,gome2022extreme,de2022bistability,de2022discontinuous,wang2023lifetimes}.

We find that the dependence of mean lifetimes on $\delta$ is consistent with an exponential relation, although a power-law dependence with a large exponent could not be ruled out. This behavior differs markedly from the parameter dependence of the lifetimes observed for 3D turbulence and LSVs in thin layers \cite{van2019rare,de2022bistability} and also from the transition to turbulence in a pipe \cite{avila2023transition}. In both of the latter cases, the timescales involved depend super-exponentially on the control parameter in question (Reynolds number in pipe flow and the ratio between forcing scale and layer height in thin-layer turbulence). More generally, the history of transitional shear flow studies suggests caution when extrapolating our findings to asymptotically long lifetimes; for instance, for turbulent puffs and slugs in pipe flow, the double-exponential dependence of lifetimes on Re is technically hard to determine, and earlier studies with less data and smaller domains incorrectly suggested an exponential dependence on Re instead \cite{hof2006finite} -- see also the discussion in \cite{avila2010transient}. 

We further find that the mean lifetimes of LSVs and jets increase approximately exponentially with the Reynolds number $\mathrm{Re}$, as the energy gap between the LSV and jet states grows with $\mathrm{Re}$. 
We also showed that the mean lifetimes increase rapidly with increasing nondimensional forcing scale $\tell$ between the forcing scale and the domain size, in agreement with previous work \cite{bouchet2014non}. This is likely the reason why the transitions described here have not been more widely observed, since 2D turbulence experiments are often performed with small-scale forcing, $\tell\ll 1$. We also observed that for a specific forcing scale, with $\tell=1/\sqrt{5}$, the lifetime of the LSV state in a square domain was drastically reduced by a spontaneous symmetry breaking, with frequent transitions between metastable LSVs and jets in both $x$ and $y$ directions. The precise quantitative origin of this phenomenon is as yet unknown. 

\avkrev{Our results are consistent with the existence of an Arrhenius-like law in this system, where the anisotropy parameter $\delta$ and the LSV-jet energy gap (measured in units of the forcing energy or large scale energy variance) control the relative and absolute lifetimes of the two states, although no theoretical argument in favor of such a law is at present available. }

We have also analyzed in detail transition trajectories between LSV and jet states, showing that the transition involves a rapid change in the flow polarity $m$ generated by efficient nonlinear energy transfer between large-scale modes, followed by a slow subsequent change in the kinetic energy $E$ as the flow amplitude adjusts to the new equilibrium state. The system trajectory in the reduced phase space spanned by $(m,E)$ forms a loop resembling relaxation oscillations, although the system traverses the loop at random times given by the lifetimes of LSVs and jets. Transition paths between multistable attractors, including turbulent flows, may be computed using the instanton formalism, cf. \cite{laurie2015computation,bouchet2019rare,grafke2019numerical,simonnet2021multistability} but whether the properties of the observed transition paths in the present system, including the two-step transition behavior revealed by the evolution of $m$ and $E$, can be derived on theoretical grounds remains an open question.

\avkrev{We also described a transition between different numbers of jets. As discussed above, viscous damping is the only mechanism for energy dissipation retained in our model. As a result there is a substantial energy gap between these states even at moderate $\R$, leading to exceedingly rare transitions. We reiterate that the choice of purely viscous condensate saturation was made to simplify the system as much as possible and to limit the number of control parameters in the system to three.} Alternative but scale-independent damping terms, such as linear Rayleigh damping \cite{boffetta2012two} or nonlinear damping (used to model turbulent drag \cite{gill1982atmosphere}), allow saturation at significantly lower kinetic energies, thereby providing benefit for numerical simulations, while also being physically relevant. These have been included in the study of zonal jets on the $\beta$-plane \cite{bouchet2019rare,cope2020dynamics}. In contrast with the well-studied problem of jets on the $\beta$-plane, however, a detailed investigation of the dynamics of purely geometry-induced jets in stochastically forced, anisotropic 2D turbulence is still outstanding, and may benefit from the inclusion of such alternative damping mechanisms to eliminate the large energy gap that makes transitions between jets so exceedingly rare.  
{Another important open problem concerning states with multiple pairs of jets in the present system is the nature of the arrest mechanism that prevents energy from accumulating in the gravest modes, given that no $\beta$ effect is present in our model. We leave these aspects of the problem for future study.}

\avkrev{Given the relatively simple transition phenomenology in terms of energy and polarity revealed by the phase space diagram shown in Fig.~\ref{fig:m_E_phase_space_portrait}, a promising avenue for a better understanding of the parameter dependence of our system might be the development of a low-dimensional reduced model. At the very least, such a reduced model needs to retain the modes at $\mathbf{k}_{1,0}$, $\mathbf{k}_{0,1}$ and the forcing-scale modes, as well as triads facilitating the interaction between the forcing scales and the large scales. }

The present study is, of course, highly idealized owing to the assumption of exact two-dimensionality and the use of periodic boundaries. \avkrev{Within these idealizations, it may be of interest to consider the related problem of 2D, stochastically forced, viscously damped flow on a sphere, varying the global rotation rate. On the non-rotating sphere, the final state produced by a stochastically forced inverse cascade is a vortex quadrupole (although this is not always observed in the unforced case due to ergodicity breaking \cite{qi2014hyperviscosity, dritschel2015late}), 
while zonal jets (and Rossby waves) emerge when the sphere is rapidly rotating \cite{qi2014hyperviscosity}. Similar results are found in the shallow water equations on the sphere \cite{cho1996emergence,scott2007forced}, but in either case, the transition between jets and vortices on the sphere remains to be analyzed. The inclusion of boundaries in two-dimensional turbulence is known to induce a rich phenomenology \cite{miller2023gyre} and it would be of interest to study the robustness of the results presented here in that setting. Another} promising direction for future work is the inclusion of the third dimension in the context of quasi-2D turbulence \cite{alexakis2023quasi}. Given the established bistability between small-scale 3D turbulence and LSVs in thin-layer turbulence near a threshold layer height \cite{de2022bistability}, there is an intriguing possibility that small-scale 3D turbulence, LSVs and large-scale jets may coexist when horizontal anisotropy is introduced in a thin 3D layer. Furthermore, it will also be important to clarify how the results described here for 2D flow are related to similar transitions between jets and LSVs in 3D turbulent rotating convection within anisotropic plane layers \cite{julien2018impact, guervilly2017jets} or on an inclined $f$-plane \cite{novi2019,barker2020,julien2022quasi,aE2023}\avkrev{, a problem which has  recently seen important theoretical advances \cite{tro2024parameterized}.} A key difference between these convection problems and our study is that buoyant momentum forcing is self-organized in response to the flow field itself, and acts over a broad range of scales, in contrast to the idealized white noise considered here. \avk{In this context, we reiterate that the longest simulations presented here extended over more than \avkrev{$10^4$ viscous time units}, a time \avkrev{that is} orders of magnitude longer than any available simulations of 3D turbulent rapidly rotating convection. For example, the runs presented in \cite{julien2018impact} only extend over $40$ viscous time units, while the longest runs reported within the rapidly rotating regime extended over $10t_\nu$ in \cite{guervilly2017jets} and $0.5t_\nu$ in \cite{favier2019subcritical}. As a result we were able to compute many more transitions and collect reliable statistical information.} \avkrev{Moreover, the white noise driving considered here and in much earlier work on 2D and quasi-2D turbulence differs from the red noise (larger noise variance at small frequencies) displayed by climatic fluctuations \cite{hasselmann1976stochastic,mitchell1976overview}, and it might be of interest to examine the effect of such colored noise on the transitions in present system as well as their potential experimental realization.} 

\section{Acknowledgment}
\avk{The authors are grateful to \avkrev{two anonymous referees for their detailed comments which have led to a significant improvement of this manuscript}, to Keith Julien and Alexis Kaminski for their comments on an earlier version of this manuscript, to Keith Julien for suggesting the remapping to a square}, \avkrev{and to Corentin Herbert for pointing out the Reynolds number dependence of the variance of the total kinetic energy}.}
L.X. was supported by the Physics Innovators Initiative (Pi\textsuperscript{2}) Summer Scholars Program, the Berkeley Physics Undergraduate Research Scholars Program (BPURS), and the Berkeley Physics Undergraduate Student Travel Scholarship in the UC Berkeley Physics Department. This work was supported by the National Science Foundation (Grants DMS-2009563, DMS-2308337, and OCE-2023541) and by the German Research Foundation (DFG Projektnummer: 522026592). The simulations described here were partly performed on the Savio computational cluster resource provided by the Berkeley Research Computing program at the University of California, Berkeley (supported by the UC Berkeley Chancellor, Vice Chancellor for Research, and Chief Information Officer). This project was also supported by the NSF ACCESS program (project number: PHY230056), allowing us to utilize the Advanced Research Computing at the Johns Hopkins (ARCH) core facility (rockfish.jhu.edu), which is supported by the National Science Foundation (NSF) grant number OAC 1920103, \avkrev{and the Purdue Anvil CPU cluster \cite{song2022anvil}.} C.L. acknowledges support from the Connecticut Sea Grant PD-23-07. 
\appendix
\begin{figure}
    \centering
    \includegraphics[width=0.8\textwidth]{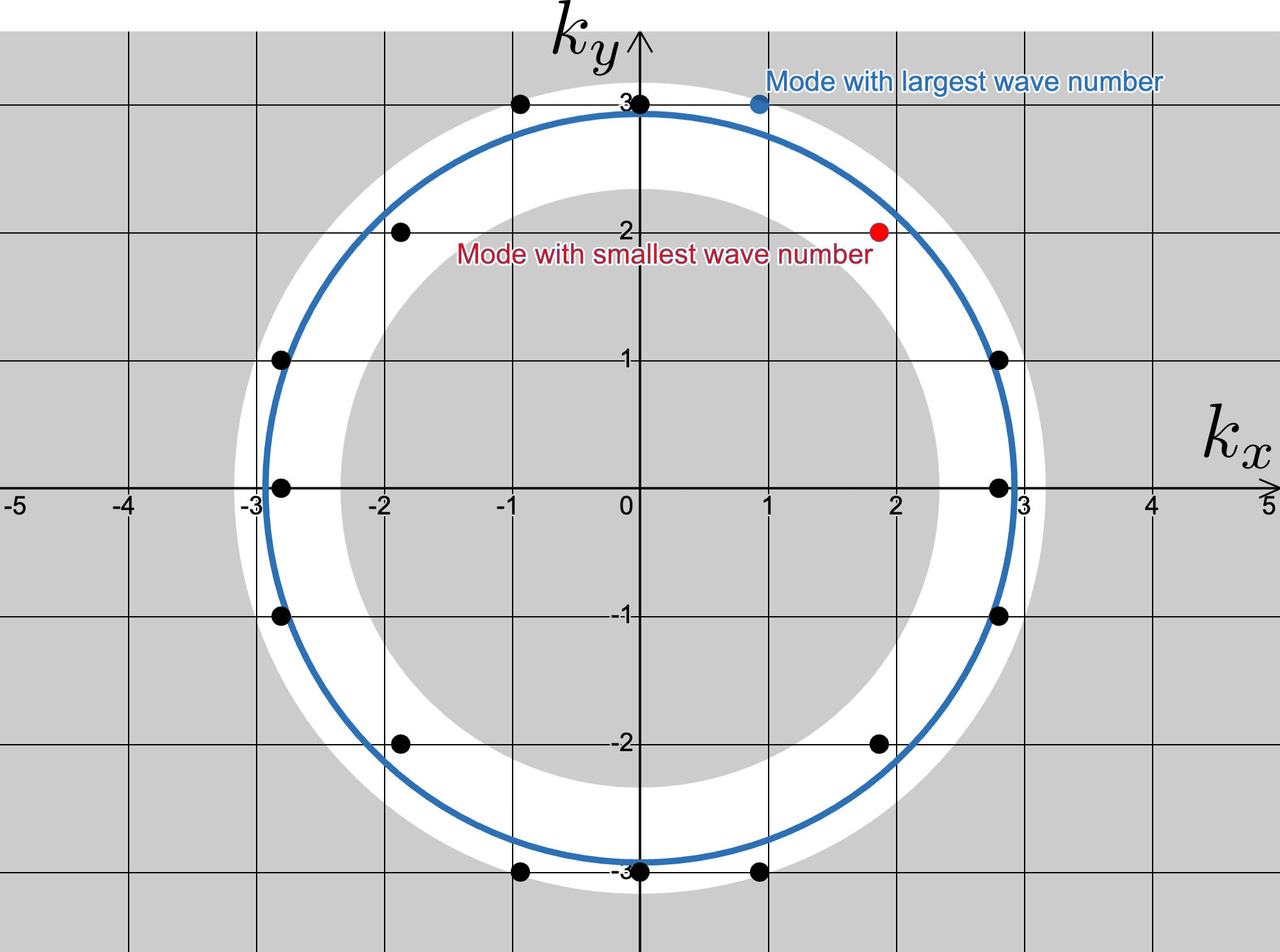}
    \caption{Illustration of the forcing band employed in all simulations in sets D1, D2, and R. In this specific case, $\delta = 1.07$.}
\label{fig:forcing_band_illustration}
\end{figure}
\section{Specification of the forcing band\label{sec:forcing_details}}
\avkrev{In a finite, anisotropic domain, the spacing of the discrete wave numbers in the $x$ and $y$ directions is not the same. Specifically, as the domain is elongated in the $x$ direction, the spacing in this direction, $Q_x=2\pi/L_x$, is reduced. For a fixed forcing range, a change of $\delta$ can change the specific forcing modes falling into the forcing range, thereby changing the degree of isotropy of the forcing and leading to visible effects in the simulation results. In our simulations, we endeavored to avoid this by an appropriate specification of the forcing band. 
For all runs in sets D1, D2 and R, which cover an interval of aspect ratios $\delta\in[1,1.1]$, we chose an appropriately narrow forcing range containing wave numbers $k\in [k_1,k_2]$, with $k_1=2.34\, Q_y$, $k_2=3.17\, Q_y$ and $Q_y=2\pi/L_y$. In this way, the forcing band contains exactly the modes $\mathbf{k}_{a,b}\equiv \left(aQ_x,bQ_y\right)$ with $(a,b)=(\pm2,2),(\pm2,-2),(\pm3,0), (0,\pm3),(\pm3,1),(\pm3,-1),(1,\pm3),(-1,\pm3)$, illustrated in Fig.~\ref{fig:forcing_band_illustration}, for all $\delta\in[1,1.1]$. 
This set of forcing modes remains close to isotropic as $\delta$ is varied. 
The dimensional forcing scale $\ell$ is taken to be the arithmetic mean of the largest and smallest scales contained in the forcing band, which is close to $0.34L_y$ for sets D1, D2, and R.}

\avkrev{
For runs in set L, we kept $\ell$ fixed at 1, while varying the domain size $L_y$ and maintaining the aspect ratio $\delta=1.07$ unchanged. 
The width of the forcing range in set L was not kept strictly constant, with the goal of maintaining the forcing as close to isotropic as possible. In set $\mathrm{J}$, we chose $k_1=1.5 Q_y$, $k_2=2.5 Q_y$. Finally, in set $\mathrm{T}$, we picked $k_1=2.12Q_y$ and $k_2=2.53Q_y$ so that only modes with wave number $\sqrt{5}$ were forced.}

\section{Remapping to the square: anisotropic diffusion and forcing\label{sec:remap_square}}
It is a simple observation that Eq.~(\ref{eq:NSE}) in the rectangular domain $(x,y)\in [0,L_x]\times[0,L_y]$ can be mapped onto to the square $(\check{x},y)\in [0,L_y]\times[0,L_y]$ by letting $x=\delta \check{x}$, implying $\partial_x = \delta^{-1} \partial_{\check{x}}$, such that the vorticity equation, written in terms of $\psi$,
\begin{equation}
    \partial_t (-\nabla^2 \psi) + J(\psi,\nabla^2 \psi) = -\nu (\nabla^2)^2 \psi  + f_\omega(x,y,t),
\end{equation}
with the Jacobian $J(a,b) = (\partial_x a)(\partial_y b) - (\partial_y a)(\partial_x b)$ and $f_\omega = \hat{z}\cdot \nabla\times \mathbf{f}$, \avk{is mapped to the following equation for the rescaled streamfunction $\check{\psi}$},
\begin{equation}
    \partial_t (-\check{\nabla}^2 \check{\psi}) + \check{J}(\check{\psi},\check{\nabla}^2 \check{\psi})= -\nu (\check{\nabla}^2)^2\check{\psi}  + \check{f}_\omega(\check{x},y,t), \label{eq:anis_hd2d}
\end{equation} 
where \avk{$\check{\psi}=\delta^{-1} \psi$}, the anisotropic Laplacian is $\check\nabla^2= \delta ^{-2} \partial_{\check{x}}^2 + \partial_y^2$ and the anisotropic Jacobian is $\check{J}(a,b) = (\partial_{\check{x}}a)(\partial_y b) - (\partial_y a) (\partial_{\check{x}} b)$. In this system the \avk{rescaled forcing $\check{f}(\check{x},y,t)=\delta^{-1}{f}_\omega(\check{x},y,t)$} drives the flow anisotropically in an ellipsoidal wave number shell, instead of the isotropic ring of wave numbers on which $f_\omega(x,y,t)$ acts. Clearly, when $\delta>1$, the dissipation in the $y$ direction is enhanced relative to the $\check{x}$-direction, a key ingredient for the transition from LSVs to jets.

Incidentally, an equation of the form (\ref{eq:anis_hd2d}) has been obtained for the barotropic vorticity in quasi-geostrophic convection on a tilted $f$-plane characterized by misaligned rotation and gravity axes \cite{julien2022quasi,aE2023}, where transitions between LSVs and jets were observed at finite inclination angles, not unlike those explored further below. We stress that in \cite{julien2022quasi,aE2023} the anisotropy is not geometrical but rather derives from the misalignment between gravity and the rotation axis.


\bibliography{references}

\end{document}